\documentclass[sn-mathphys,iicol]{sn-jnl}

\usepackage{bm}
\usepackage{xcolor}
\usepackage{listings}
\usepackage{xurl}
\usepackage{caption}
\usepackage{footnote}
\makesavenoteenv{tabular}
\usepackage{subcaption}
\usepackage{etoolbox}
\makeatletter
\patchcmd{\ps@headings}
{\hbox to \hsize{\hfill Springer Nature 2021 \LaTeX\ template\hfill}}
{\hbox to \hsize{\hfill \hfill}}
{}
{}
\patchcmd{\ps@titlepage}
{\hbox to \hsize{\hfill Springer Nature 2021 \LaTeX\ template\hfill}}
{\hbox to \hsize{\hfill \hfill}}
{}
{}
\patchcmd{\ps@reference}
{\hbox to \hsize{\hfill Springer Nature 2021 \LaTeX\ template\hfill}}
{\hbox to \hsize{\hfill \hfill}}
{}
{}
\makeatother

\newcommand{\cps}{CPS}

\newcommand{\rtamt}{RTAMT}
\newcommand{\rosrtamt}{RTAMT4ROS}
\newcommand{\antlr}{ANTLR4}
\newcommand{\AST}{AST}

\newcommand{\ros}{ROS}
\newcommand{\hsr}{HSR}

\newcommand{\matlabSimulink}{MATLAB/Simulink}
\newcommand{\python}{Python}
\newcommand{\cpp}{C++}

\newcommand{\circelNum}[1]{\raise0.2ex\hbox{\textcircled{\scriptsize{#1}}}}

\newcommand{\Reals}{\mathbb{R}}
\newcommand{\Naturals}{\mathbb{N}}
\newcommand{\Times}{\mathbb{T}}

\newcommand{\ground}[1]{{#1}^{*}}

\newcommand{\pos}{\mathbf{x}}
\newcommand{\agentpos}[1]{\pos_{\agent{#1}}}
\newcommand{\agent}[1]{\mathbf{a}_{#1}}
\newcommand{\agents}{\mathcal{A}}
\newcommand{\obstacle}[1]{O_{#1}}
\newcommand{\obstacles}{\mathcal{O}}
\newcommand{\prohibregion}[1]{B_{#1}}
\newcommand{\prohibregions}{\mathcal{B}}

\newcommand{\Locations}{\mathcal{L}}
\newcommand{\waypoint}[1]{\Locations_{#1}}
\newcommand{\waypoints}{\mathcal{W}}

\newcommand{\Dist}{\mathbf{Dist}}
\newcommand{\errfn}{\mathit{err}}

\newcommand{\ltl}{LTL}
\newcommand{\stl}{STL}
\newcommand{\bfstl}{bfSTL}
\newcommand{\pstl}{pSTL}
\newcommand{\iastl}{IA-STL}

\newcommand{\timeElement}{t}

\newcommand{\valueElement}{v}

\newcommand{\signal}{w}

\newcommand{\interval}{I}

\newcommand{\fun}{\varphi}
\newcommand{\rob}{\rho}

\newcommand{\imply}{\to}

\newcommand{\until}{\,\mathbf{U}\,}
\newcommand{\globally}{\mathbf{G}\,}
\newcommand{\finally}{\mathbf{F}\,}
\newcommand{\Next}{\mathbf{X}\,}

\newcommand{\since}{\,\mathbf{S}\,}
\newcommand{\once}{\mathbf{O}\,}
\newcommand{\historically}{\mathbf{H}\,}
\newcommand{\previous}{\mathbf{Y}\,}
\newcommand{\precedes}{\mathbf{P}\,}

\newcommand{\horizon}{h}
\newcommand{\temporalDepth}{H}
\newcommand{\pastificationOperation}{\Pi}

\jyear{2021}%

\theoremstyle{thmstyleone}%
\theoremstyle{thmstyletwo}%
\newtheorem{example}{Example}%

\theoremstyle{thmstylethree}%
\newtheorem{definition}{Definition}%

\definecolor{codegreen}{rgb}{0,0.6,0}
\definecolor{codegray}{rgb}{0.5,0.5,0.5}
\definecolor{codepurple}{rgb}{0.58,0,0.82}
\definecolor{backcolour}{rgb}{0.95,0.95,0.92}

\lstdefinestyle{mystyle}{
    language=Python,
    backgroundcolor=\color{backcolour},   
    commentstyle=\color{codegreen},
    keywordstyle=\color{magenta},
    numberstyle=\tiny\color{codegray},
    stringstyle=\color{codepurple},
    basicstyle=\ttfamily\footnotesize,
    breakatwhitespace=false,
    breaklines=true,
    captionpos=b,
    keepspaces=true,
    numbers=left,
    numbersep=5pt,
    showspaces=false,
    showstringspaces=false,
    showtabs=false,
    tabsize=2
}

\begin{document}

\title[\rtamt]{\rtamt - Runtime Robustness Monitors with Application to CPS and Robotics}


\author[1]{\fnm{Tomoya} \sur{Yamaguchi}}\email{tomoya.yamaguchi@toyota.com}
\author[1]{\fnm{Bardh} \sur{Hoxha}}\email{bardh.hoxha@toyota.com}
\author[2]{\fnm{Dejan} \sur{Ni\v{c}kovi\'{c}}}\email{dejan.nickovic@ait.ac.at}

\affil[1]{\orgdiv{TRINA}, \orgname{Toyota Motor NA R\&D}, \orgaddress{\street{1555 Woodridge Ave}, \city{Ann Arbor}, \postcode{48105}, \state{Michigan}, \country{U.S}}}
\affil[2]{\orgname{AIT Austrian Institute of Technology}, \orgaddress{\street{Giefinggasse 4},\city{Vienna}, \postcode{1210}, \state{Vienna}, \country{Austria}}}


\abstract{
    In this paper, we present Real-Time Analog Monitoring Tool (\rtamt), a tool for quantitative monitoring of Signal Temporal Logic (\stl) specifications.
The library implements a flexible architecture that supports: 
(1) various environments connected by an Application Programming Interface (API) in \python,
(2) various flavors of temporal logic specification and robustness notion such as \stl, including an interface-aware variant that distinguishes between input and output variables, and
(3) discrete-time and dense-time interpretation of \stl\ with generation of online and offline monitors.
We specifically focus on robotics and Cyber-Physical Systems (\cps s) applications, showing how to integrate \rtamt\ with
(1) the Robot Operating System (\ros) and 
(2) \matlabSimulink\ environments. 
We evaluate the tool by demonstrating several use scenarios involving service robotic and avionic applications.
}




\maketitle

\section{Introduction}
\label{sec:intro}

\begin{figure*}[t]
    \centering
    \begin{minipage}{0.4\hsize}
        \centering
        \includegraphics[width=\textwidth]{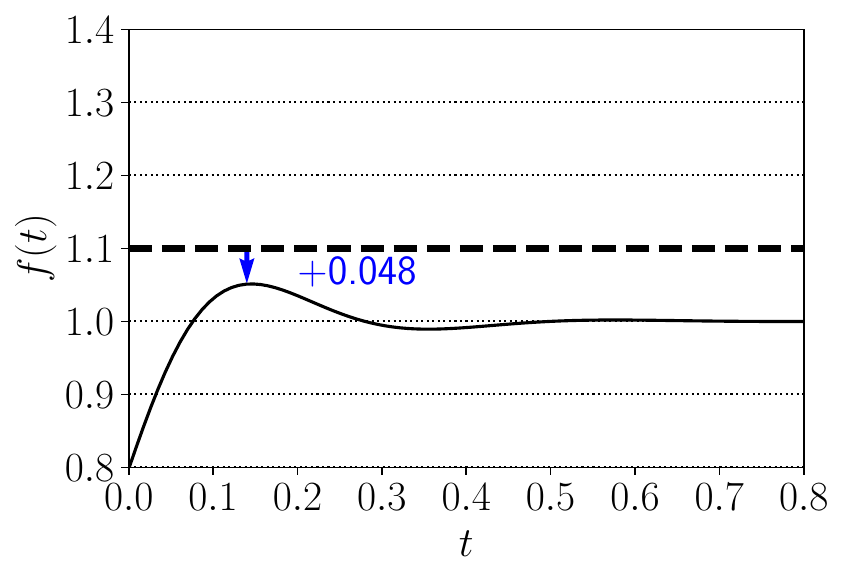}
        \subcaption{Satisfied}
    \end{minipage}
    \hspace{24pt}
    \begin{minipage}{0.375\hsize}
        \centering
        \includegraphics[width=\textwidth]{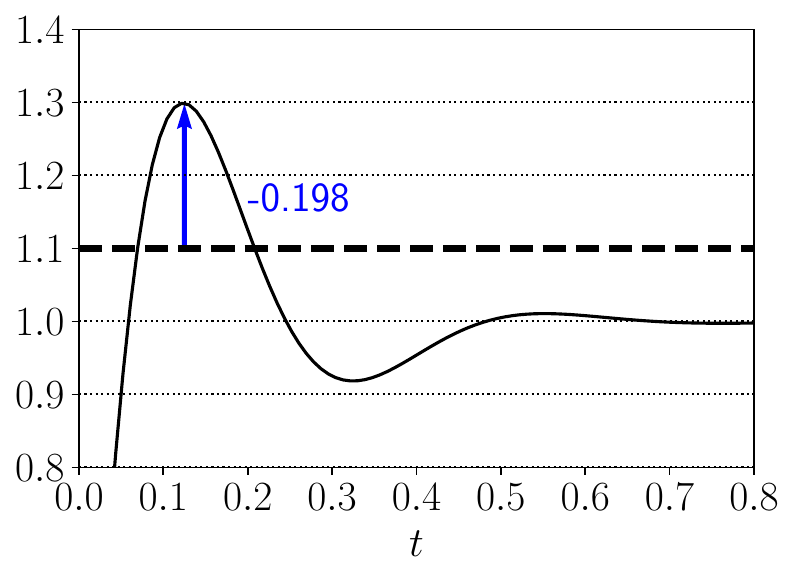}
        \subcaption{Violated}
    \end{minipage}
    \caption{Two typical PID controller behaviors evaluated against the specification "it is always the case that $f(t)$ is smaller or equal to 1.1" ($\globally(f(t) \leq 1.1)$) in \stl: apart from the satisfied/violated qualitative verdict, the robustness semantics provides additional quantitative feedback  (a) satisfied with $robustness=+0.048$ (b) violated with $robustness=-0.198$.}
    \label{fig:robustness}
\end{figure*}

Cyber-Physical Systems (\cps s)~\cite{lee2016introduction, mitra2021verifying, alur2015principles} are systems which integrate both cyber and physical components, operating in complex environments and increasingly featuring autonomous decision-making capabilities enabled by machine learning.
Distributed robotic applications are typical examples of autonomous \cps\ used in contexts ranging from collaborative manufacturing environments to assisted living. 
Other examples of \cps s include smart buildings that learn their practitioner's profile and accordingly optimize heating strategies, autonomous vehicles that are able 
to drive without human intervention or medical devices that modify a therapy according to the patient's needs.
To address complexity challenges, various frameworks have been proposed to facilitate development. 
For instance, Robot Operating System (\ros)~\cite{ros} provides a meta-operating system with tools and libraries to help engineers develop robotic applications while \matlabSimulink\textsuperscript{\textregistered} enables the modeling of \cps\ control applications.

Under these circumstances, Verification and Validation (V\&V) remains a bottleneck as state-of-the-art techniques do not scale to this level of complexity, making static safety assurance a very costly and time-demanding, if not impossible, activity.
Runtime Assurance (RTA)~\cite{sha2001using}, an alternative approach for ensuring the safe operation of robotic and other sophisticated \cps s, is used when static verification is not possible.
RTA allows the use of unverified components in a system that implements a safe fallback mechanism for 
(1) detecting anomalies during real-time system operations and 
(2) invoking a recovery mechanism that can bring the system back to its safe (and possibly degraded) operation.
Runtime verification provides a reliable, rigorous and systematic way for finding violation in system executions and consequently represents a viable solution for the monitoring part of an RTA entity.
Runtime verification and assurance techniques must support seamless integration in common design and operation environments.

Formal specifications play an important role in runtime verification and enable to precisely express intended system properties.
Signal Temporal Logic (\stl)~\cite{mn04} is a formal specification language that is used to describe \cps\ requirements.
\stl\ can provide quantitative \emph{robustness} semantics, which is used to measure to which extent an observed behavior satisfies or violates a given specification. For instance, Fig.~\ref{fig:robustness} shows two typical PID controller behaviors. The STL specification requires the observed behavior $f(t)$ to always remain below $1.1$. The first behavior (Fig.~\ref{fig:robustness}~(a)) satisfies its specification with the positive robustness $+0.048$. This value corresponds to the closest distance between $f(t)$ and the $1.1$ threshold. In contrast, the second behavior (Fig.~\ref{fig:robustness}~(b)) violates that same specification with negative robustness $-0.198$. This value represents the largest amount that $f(t)$ goes above $1.1$.   
We observe that the real-valued evaluation contrasts the classical satisfied/violated answer that we typically get from reasoning with the qualitative interpretation of specification languages.

To address the above V\&V challenges, we present Real-Time Analog Monitoring Tool (\rtamt)\footnote{\rtamt:\url{https://github.com/nickovic/rtamt}}, a versatile library for generating monitors from \stl\ specifications.
The contributions of this work are as follows:
\begin{enumerate}
    \item We integrate \rtamt\ into multiple design and operational environments with its Application Programming Interface (API) -- the Python library's API allows for this integration.
    We demonstrate the integration of \rtamt\ with \ros\footnote{\rosrtamt:\url{https://github.com/nickovic/rtamt4ros}} and \matlabSimulink.
    \item The tool facilitates temporal logic-based specifications and quantitative/robustness notions such as \stl.
    We allow integration of various syntactic and semantic variants of the language.
    For instance, the tool supports standard \stl\ and its Interface-Aware extension (\iastl)~\cite{iastl}.
    We also provide a tool library to support implementing one's own temporal logic-based specification language.
    \item We implement both \emph{offline} and \emph{online} monitors with either \emph{discrete-time} or \emph{dense-time} interpretation of the behaviors and the specification language.
\end{enumerate}

This work is an extended version of the conference paper~\cite{rtamt}. We briefly summarize the additional content that is provided in this paper, compared to its conference variant:
\begin{itemize}
    \item A detailed presentation of the library's architecture.
    \item An extensive description of how \rtamt\ can be integrated to \ros\ and \matlabSimulink\textsuperscript{\textregistered}.
    \item A comprehensive evaluation section with additional experiments and case studies.
    \item More extensive and complete related work section.
\end{itemize}

The layout of the paper is as follows.
First, we present definitions of \stl\ and its extensions and sketch the monitoring procedures in Sec.~\ref{sec:tempralLogic}; we then give the \rtamt\ architecture in Sec.~\ref{sec:architecture}, its API in Sec.~\ref{sec:api}, and its library in Sec.~\ref{sec:library}.
These enable the use of specification-based runtime verification and assurance methods.
In Sec.~\ref{sec:experiments}, we give our detailed evaluation of the tool and case studies in robotic applications based on \ros\ and \matlabSimulink. We then present related work in Sec.~\ref{sec:relatedWork}, and finally conclude the paper in Sec.~\ref{sec:conclusion}.

\section{Monitoring Temporal Logic Specifications}
\label{sec:tempralLogic}

\rtamt\ is a tool for the automatic generation of monitors from declarative specifications.
Given an input signal in the form of a sequence of (time, value) pairs and a specification, \rtamt\ computes at different points in time how robust the observed signal is compared to the specification, i.e. how far is it from satisfying or violating it. 
\stl\ is utilized as the  specification language of choice in \rtamt, which supports discrete-time and dense-time monitors for usage in a wide array of applications. 

In Sec. \ref{sec:stl}, we provide an overview of the syntax of \stl\ with the standard \emph{future temporal operators} and more specialized \emph{past temporal operators}, which are more suitable for online monitoring requirements.
With a specification that includes only past temporal operators, the robustness of a signal with respect to the specification may be evaluated at the current timestep and does not depend on future observations.
In addition, we provide semantics that support discrete and dense temporal models. 
In Sec. \ref{sec:pastification} we present a pastification procedure that enables translation of future temporal operators to past temporal operators.
Finally, in Sec.~\ref{sec:iastl} we introduce the \emph{interface-aware} extension of \stl\ (\iastl) that distinguishes between \emph{input} and \emph{output} variables.
We use \iastl\ to demonstrate how the library can be extended with other specification formalisms such as temporal operators and additional robustness metrics.

\subsection{Signal Temporal Logic}
\label{sec:stl}

\begin{figure*}[t]
   \centering
   \begin{minipage}{0.48\hsize}
        \centering
        \includegraphics[width=\linewidth]{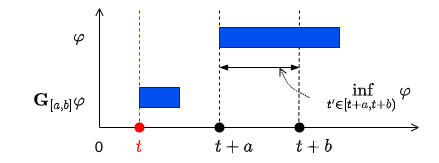}
        \subcaption{$\rob( \globally{[a,b]}\,\fun, \signal, \timeElement) = \displaystyle\inf_{\timeElement' \in [\timeElement+a,\timeElement+b)} \rob(\fun, \signal, \timeElement')$}
        \label{fig:globally}
    \end{minipage}
    \begin{minipage}{0.48\hsize}
        \centering
        \includegraphics[width=\linewidth]{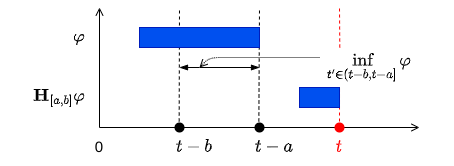}
        \subcaption{$\rob( \historically_{[a,b]}\,\fun, \signal, t) = \displaystyle\inf_{\timeElement' \in (\timeElement-a,\timeElement-b]} \rob(\fun, \signal, \timeElement')$}
        \label{fig:historically}
    \end{minipage}
    \caption{Semantics for the future timed operator $\globally$ (globally) and the past timed operator $\historically$ (historically): Input $\fun$ is described as Booleans for simplicity.
    The difference is 
    (a) $\timeElement'\in [\timeElement+a, \timeElement+b]$ in $\globally_{[a,b]}$ refers to future points in time with respect to the current time $\timeElement$, 
    In contrast, 
    (b) $\timeElement'\in [\timeElement-a, \timeElement-b]$ in $\historically_{[a,b]}$ refers to past points in time with respect to the current time $\timeElement$,}
    \label{fig:globallyHistorically}
\end{figure*}

Signal Temporal Logic (\stl) extends Linear Temporal Logic (\ltl) with real-time temporal operators and numerical predicates defined over real-valued behaviors. 
We interpret \stl\ over \emph{finite signals} that we represent as finite sequences of (time, value) pairs. Let $X = \{ x_1, \ldots, x_n \}$ be a set of real-valued variables.
A valuation $v : X \rightarrow \Reals$ For $x \in X$ maps a variable $x$ to a real value. 
A signal $\signal$ defined over $X$ is a function $\Times \to \Reals^X$ that gives the value $\signal(t)$ of the variables in 
$X$ at time $t \in \Times$, where 
$\Times$ is a finite interval $[0,d)$. A signal $w$ can be also seen as a vector of real-valued signals $w_x~:~\Times \to \Reals$ associated to variables $x \in X$. 
Given two signals $w^{1}~:~\Times \to \Reals^{n_1}$ and $w^{2}~:~\Times \to \Reals^{n_2}$, we define their composition $w^{1} \mid\mid w^{2}$ as the function $\Times \to \Reals^{n_1 + n_2}$, with the expected meaning. Given a subset $Y \subseteq X$ of variables, we define the projection $w_Y$ of the signal $w$ to $Y$ as the composition $\mid\mid_{y \in Y} w_y$.

We can have two interpretations of signals -- discrete- and dense-time. 
For the discrete-time interpretation of signals, we assume that the signal is periodically sampled with some period $\Delta$. 
It follows that every sample can be uniquely indexed with an integer $i$ corresponding to the sampling time $i\Delta$. Consequently, the time domain $\Times = [0,d]$ is an interval of integers with $d \in \mathbb{N}$. 
For the dense-time interpretation, we assume that the time domain $\Times=[0,d]$ is an interval of reals with $d \in \mathbb{Q}_{> 0}$.
In contrast to the discrete-time interpretation, the changes in the dense-time signal can happen anywhere in the interval $[0,d]$.
While we assume signals with finite variability, we do not impose a bound on the number of changes in any unit time.

Both discrete-time and dense-time signals can be represented with a finite sequence $(\timeElement_0,\valueElement_0),\ldots,(\timeElement_n,\valueElement_n)$ of (timestamp, valuation) pairs such that $\timeElement_0=0$ and $\forall i \in[1,n] : t_i \in \Times$ and $\timeElement_{i-1}<\timeElement_i$,  $\valueElement_i$ is a valuation over $X$ and $t_n = d$.
For the dense-time interpretation of the specification language, we assume that signals are piecewise-constant, i.e. for all $t \in [t_i, t_{i+1})$ and $x \in X$, $w(x,t) = w(x,t_i)$.



The basic building block in \stl\ is a \emph{predicate} $f(Y)>c$, where $f(Y)$ is a term with $Y \subseteq X$, $f~:~\Reals^Y \to \Reals$ is an interpreted function and $c$ is a real number.
The syntax of an \stl\ formula $\varphi$ is defined recursively with the following grammar,

\begin{definition}[Syntax of \stl]
$$
\begin{array}{lcl}
    \fun & := & f(Y)>c \mid
    \neg \fun \mid
    \fun_1 \vee \fun_2 \mid  \fun_1 \until_\interval \fun_2 \mid
    \fun_1 \since_\interval \fun_2 \\
\end{array}
$$
\noindent where $Y \subseteq X$, $\interval$ is an interval of the form $[a,b]$ or $[a, \infty)$ where $a \leq b$ are rational numbers.
\end{definition}

The operators for disjunction $(\vee)$ and negation $(\neg)$ are defined in the usual way. 
$\until$ (until) and $\since$ (since) are temporal operators.
The syntax of \stl\ is very similar to that of \ltl, with the addition of numerical predicates of the form $f(Y)>c$ and intervals $\interval$ that bound the scope of the temporal operators.


Given an STL formula $\varphi$, a signal $w$ and a time $t \in \mathbb{T}$, we define the \emph{quantitative} (or \emph{robustness}) semantics $\rho(\varphi, w, t)$ as follows:

\begin{definition}[Quantitative Semantics of \stl]
\label{def:semanticsSince}
\begin{align}
&\rob( f(Y) > c, \signal, \timeElement) &=& f(w_Y(t))-c \notag\\
&\rob( \lnot \fun, \signal, \timeElement) &=& -\rob(\fun, \signal, \timeElement) \notag\\
&\rob( \fun_1 \vee \fun_2, \signal, \timeElement) &=& \max\left(\rob(\fun_1, \signal, \timeElement), \rob(\fun_2, \signal, \timeElement)\right) \notag
\end{align}
\begin{align}
&\rob( \fun_1 \until_{I} \fun_2, \signal, \timeElement) = \notag\\
&\displaystyle\sup_{\timeElement' \in \timeElement \oplus I \cap \Times} 
    \min\left(
        \rob(\fun_2, \signal, \timeElement'),
        \displaystyle\inf_{\timeElement'' \in [\timeElement,\timeElement')} \rob(\fun_1, \signal, \timeElement'') 
    \right) \notag \\
&\rob( \fun_1 \since_{I} \fun_2, \signal, \timeElement) = \notag\\
&\displaystyle\sup_{\timeElement' \in \timeElement' \ominus I \cap \Times} 
    \min\left(
        \rob(\fun_2, \signal, \timeElement'),
        \displaystyle\inf_{\timeElement'' \in (\timeElement',\timeElement]} \rob(\fun_1, \signal, \timeElement'') 
    \right) \notag
\end{align}
\noindent where $\min(P)$ and $\max(P)$ denote the smallest and the largest elements in the set $P$, while $\inf(P)$ and $\sup(P)$ denote the infinum (greatest lower bound) and the supremum (least upper bound) of $P$.
\end{definition}


The quantitative semantics from Definition~\ref{def:semanticsSince}, intuitively measures how much should the signal be modified in order to satisfy or violate the specification. The measured value is also known as \emph{spatial robustness}. It can be seen as (an approximation of)  the distance between the observed behavior and the boundary of the set representing all behaviors that satisfy the property. 

We observe that the $\until$ (Until) and $\since$ (Since) operators may be utilized to derive other temporal operators:\\
Future Temporal Operators
$$
\begin{array}{llcl}
    \text{finally} & \finally_\interval \fun & \equiv & \true \until_\interval \fun \\
    \text{globally} & \globally_\interval \fun & \equiv & \neg \finally_\interval \neg \fun \\
    \text{next} & \Next \fun & \equiv & \false \until_{[0,\infty)} \fun
\end{array}
$$
Past Temporal Operators
$$
\begin{array}{llcl}
    \text{once} & \once_I \fun & \equiv & \true \since_I \fun\\
    \text{historically} & \historically_I \fun & \equiv & \neg \once_I \neg \fun \\ 
    \text{previous} & \previous \fun & \equiv & \false \since_{[0,\infty)} \fun \\
    \text{rise} & \uparrow \fun & \equiv & \previous \neg \fun \wedge \fun \\
    \text{fall} & \downarrow \fun & \equiv & \previous \fun \wedge \neg \fun
\end{array}
$$

Intuitively, a signal $\signal_Y(t)$ satisfies a formula $\fun_1 \until_{[a,b]} \fun_2$ at time $t$ if there exists a time $t' \in [t+a,t+b]$ such that $\signal$ satisfies $\fun_2$ and for all the times before then $\signal$ satisfies $\fun_1$.
Since the robustness evaluation needs to consider the signal at future timesteps, $\until$ and derived operators $\finally$, $\globally$, $\Next$ are categorized as future operators. 
Similarly, a signal $\signal_Y(t)$ satisfies a formula $\fun_1 \since_{[a,b]} \fun_2$ at time $t$ if there exists a time $t' \in [t-a,t-b]$ such that $\signal$ satisfies $\fun_2$ and for all the times after then $\signal$ satisfies $\fun_1$. 
$\since$ and derived operators $\once$, $\historically$, $\previous$ are referred to as past operators. 
The intuition behind these operators is presented visually in Figs. \ref{fig:globally} and \ref{fig:historically}. We note that discrete-time quantitative semantics of \stl\ are evaluated only at sampled instances.
Past temporal operators are more suitable for monitoring applications than future temporal operators because robustness evaluation relies only on current and past time valuations of a signal.

We note that Definition~\ref{def:semanticsSince} can 
be used for both the discrete and dense time interpretation of the logic, depending on whether $t$, $t'$ and $t''$ 
are quantified over naturals or reals. We also observe 
that the next $\Next$, previous $\previous$, rise $\uparrow$ and fall $\downarrow$ operators are only 
meaningful for the discrete-time interpretation of \stl.

In the following, we consider two subsets of \stl:
\begin{itemize}
    \item \emph{Bounded-Future} Signal Temporal Logic (\bfstl) where the unbounded time intervals are not permitted. That is, the  $\until{}_\interval$ operator has a defined interval $[a,b]$ where $a$ and $b$ are in $\Reals^+$.
    \item \emph{Past} Signal Temporal Logic (\pstl) that only supports past-operators.
\end{itemize}

To facilitate the online monitoring of bounded future properties in \rtamt, we use a \emph{pastification} procedure, which takes a \bfstl\ formula and generates an equi-satisfiable \pstl\ formula for monitoring purposes.

\subsection{Pastification from \bfstl\ to \pstl}
\label{sec:pastification}
Monitoring specifications with future temporal operators is challenging because the evaluation at time index $\timeElement$ may depend on the observed inputs at some future time indices $\timeElement' > \timeElement$.
\bfstl\ specifications have a bounded future horizon $\horizon$ that can be syntactically computed from the formula structure.
For such specifications, the online monitoring challenge can be addressed by postponing the formula evaluation from time index $\timeElement$ to the end of the (bounded) horizon $\timeElement+\horizon$, where all the inputs necessary for computing the robustness degree are available. 
In this section, we briefly sketch this procedure, called {\em pastification}~\cite{bounded,ptstl}.

We first define the {\em temporal depth} $\temporalDepth(\fun)$ of $\fun$ as the syntax-dependent upper bound on the actual depth of the specification. It corresponds to the maximum time in the future that is relevant for the evaluation of the specification now, which is inductively computed as follows\footnote{The pastification of the $\Next$ operator is relevant only for the discrete-time interpretation of the specification language.}:
\begin{definition}[Temporal Depth]
$$
\begin{array}{lcl}
    \temporalDepth(f(Y) > c) & = & 0 \\
    \temporalDepth(\neg \fun) & = & \temporalDepth(\fun) \\
    \temporalDepth(\fun_1 \vee \fun_2) & = & \max \{ \temporalDepth(\fun_1), \temporalDepth(\fun_2) \} \\
    \temporalDepth(\Next \fun) & = & \temporalDepth(\fun) + 1\\
    \temporalDepth(\fun_1 \until_{[a, b]} \fun_2) & = & b + \max \{\temporalDepth(\fun_1), \temporalDepth(\fun_2) \}
\end{array}
$$ 
\end{definition}

In the next step, we define the pastification operation $\pastificationOperation$ on the $\stl$ formula $\fun$ with past and bounded future and its bounded horizon $d = \temporalDepth(\fun)$. To enable this transformation of 
specifications, we define a new \emph{precedes}  $\precedes_{[a,b]}$ 
auxiliary temporal operator, which is essentially implementing the bounded until operator interpreted from the end of the bounded formula horizon:
\begin{align*}
&\rob( \fun_1 \precedes_{[a,b]} \fun_2, \signal, \timeElement) =\notag\\
&\displaystyle\sup_{\timeElement' \in [\timeElement-b+a,\timeElement]} 
    \min\left(
        \rob(\fun_2, \signal, \timeElement'),
        \displaystyle\inf_{\timeElement'' \in (\timeElement -b,\timeElement']} \rob(\fun_1, \signal, \timeElement'') 
    \right)
\end{align*}


The pastification procedure essentially takes a bounded future formula $\varphi$ with horizon $H(\varphi)$ and gives the recipe how to shift the evaluation of that formula from time $t$ to time $t + H(\varphi)$, when all the information required for the evaluation of the formula becomes available.

\begin{definition}[Pastification Operation]
$$
\begin{array}{lcl}
    \pastificationOperation(f(Y) > c, d) & = & \once_{[d,d]} (f(Y) > c) \\
    \pastificationOperation(\neg \fun, d) & = & \neg \pastificationOperation(\fun, d) \\
    \pastificationOperation(\fun_1 \vee \fun_2, d) & = & \pastificationOperation(\fun_1, d) \vee \pastificationOperation(\fun_2, d) \\
    \pastificationOperation(\Next \fun, d) & = & \pastificationOperation(\fun, d-1) \\
    \pastificationOperation(\finally_{[a,b]} \fun, d) & = & \once_{[0,b-a]} \pastificationOperation(\fun, d-b) \\
    \pastificationOperation(\fun_1 \until_{[a,b]} \fun_2, d) 
    &\leftrightarrow& \pastificationOperation(\fun_1, d-b) 
    \precedes_{[a,b]} \pastificationOperation(\fun_2, d-b) \\
\end{array}
$$
\end{definition}

Formally, we say that for an arbitrary \bfstl\ formula $\fun$, signal $\signal$ and time index $\timeElement \in \Naturals$, $\rho(\fun, \signal, \timeElement) = \rho(\pastificationOperation(\fun), \signal, \horizon(\fun))$.

\begin{example}
The pastification of the \bfstl\ specification $\fun \equiv (req \geq 3) \imply \finally_{[0,5]} (gnt \geq 3)$ from the running example corresponds to the \pstl formula $\Pi(\fun) \equiv \once_{[5,5]} (req \geq 3) \imply \once_{[0,5]} (gnt \geq 3)$.
\end{example}

\subsection{Interface-Aware \stl}
\label{sec:iastl}

\begin{figure*}[t]
   \centering
   \begin{minipage}{0.4\hsize}
       \includegraphics[width=\linewidth]{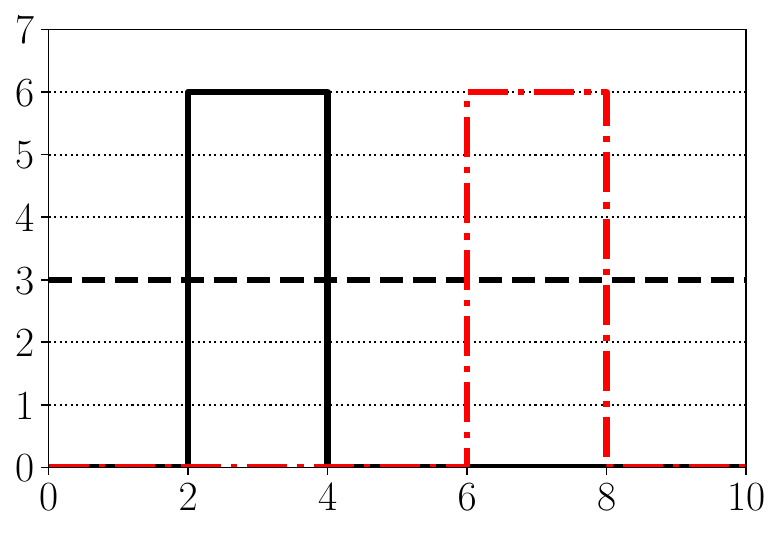}
       \subcaption{}
       \label{fig:exampleIA_a}
   \end{minipage}
   \hspace{24pt}
   \begin{minipage}{0.4\hsize}
       \includegraphics[width=\linewidth]{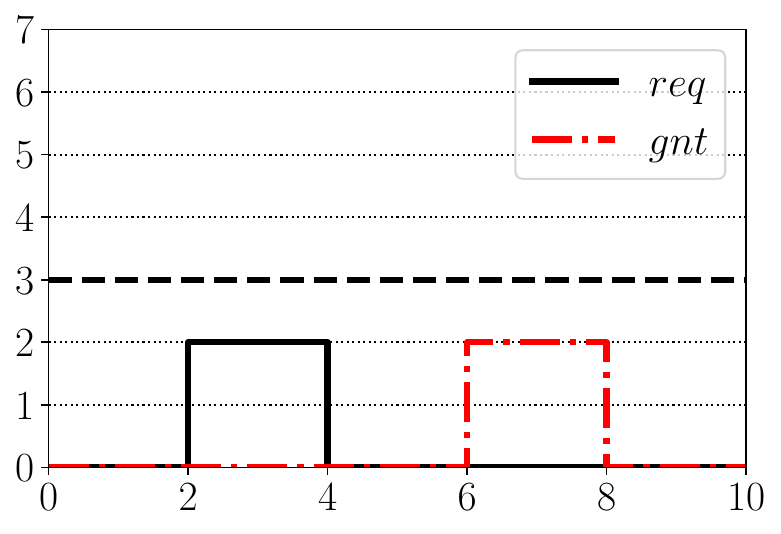}
       \subcaption{}
       \label{fig:exampleIA_b}
   \end{minipage}
   \\
   \begin{minipage}{0.4\hsize}
       \includegraphics[width=\linewidth]{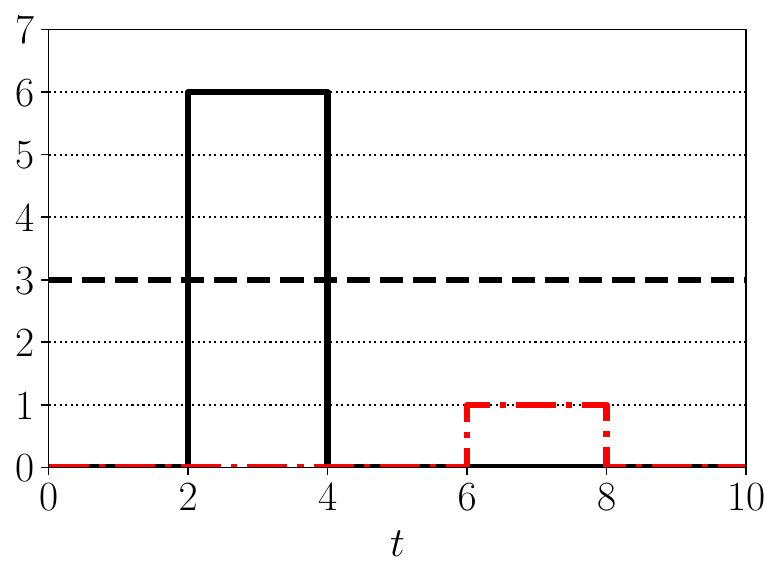}
       \subcaption{}
       \label{fig:exampleIA_c}
   \end{minipage}
   \hspace{24pt}
   \begin{minipage}{0.4\hsize}
       \includegraphics[width=\linewidth]{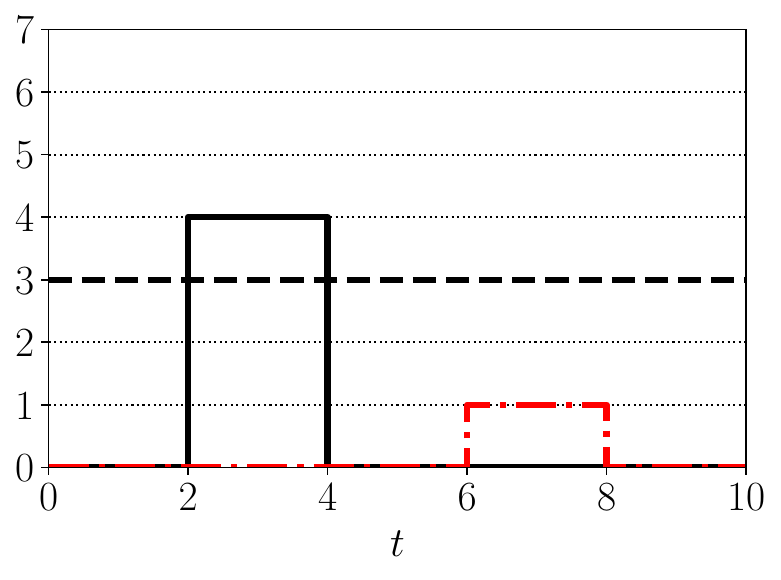}
       \subcaption{}
       \label{fig:exampleIA_d}
   \end{minipage}
   \caption{Examples of \iastl\ evaluating request-grant signals with the formula $\globally(req \geq 3 \imply \finally_{[0,5]} gnt \geq 3)$: A request is issued whenever the value of the signal $req$ is greater or equal to $3$. 
   Similarly, the grant is issued whenever the value of the signal $gnt$ is greater or equal to $3$.
   The specification requires that every request is eventually followed within $5$ time units by a grant. 
   The sub-figures illustrate $4$ different behaviors where the specification is 
   (a) satisfied with with robustness $3$, output robustness $+3$, and input vacuity $0$,
   (b) vacuously satisfied with that (the request is never issued) with robustness +1, output robustness $+\infty$, and input vacuity $+1$,
   (c) does not satisfy that with robustness $-2$ and output robustness $-2$, 
   (d) also does not satisfy that with robustness $-1$ and output robustness $-2$.}
   \label{fig:exampleIA}
\end{figure*}

Interface-aware Signal Temporal Logic (\iastl) extends \stl\ by classifying variables appearing in the specification as \emph{input} or \emph{output} variables \cite{iastl}.
This simple addition to the specification language is fundamental to reasoning about \emph{open} systems and allows their specification as input/output relations, rather than sets of correct execution traces. 
\iastl\ allows questions that cannot be formulated with the general \stl\ formulas:
\begin{itemize}
    \item How good is the reaction of the system to a given input signal 
    with respect to its requirements?
    \item Does a concrete input signal exercise the system in any meaningful way with respect to its requirements?
\end{itemize}

\iastl\ admits several semantic interpretations, one for each question that the interpretation shall answer.
There are two particularly useful \iastl\ semantics:
\begin{description}
    \item[Output robustness] measures how robust a specification is relative to the set output signals.
    The real-valued (positive or negative) robustness value (positive or negative) indicates how much the signal can be perturbed and still satisfy/violate the specification.
    When (plus or minus) infinity, the output robustness indicates that the specification is vacuously satisfied or not satisfied by the given input signal.
    \item[Input vacuity] represents the level of vacuity of the specification with respect to a given input signal. 
    When a positive or negative real, it indicates how much one can change the input without violating or satisfying the specification. 
    When equal to $0$, the input vacuity indicates the specification is non-vacuously exercised by the input signal. Intuitively, it means that the specification is not robust to even a slightest change in the inputs because it might induce an output behavior that affects the robustness in an arbitrary manner.
\end{description}

We illustrate these two notions of robustness in Fig.~\ref{fig:exampleIA}, which depicts four simple request-grant behaviors that we evaluate against the \iastl\ bounded response property $\globally(req \geq 3 \imply \finally_{[0,5]} gnt \geq 3)$.
The first behavior (Fig.~\ref{fig:exampleIA_a}) satisfies the specification with output robustness $+3$ because the $gnt$ signal is $3$ units away from the threshold in the interval of interest. 
Since the specification is not vacuously satisfied by the input $req$, the input vacuity equals to $0$. 
The second behavior (Fig.~\ref{fig:exampleIA_b}) satisfies the specification with robustness $+1$, but with output robustness $+\infty$ because the request is never issued and hence the specification is vacuously satisfied. 
The input vacuity also equals to $+1$ because changing the values of the input $req$ by any amount smaller than $1$ would still guarantee vacuous satisfaction of the specification. 
The third behavior (Fig.~\ref{fig:exampleIA_c}) violates the specification with output robustness $-2$, corresponding to the distance of $gnt$ from the threshold. 
Finally, the fourth behavior (Fig.~\ref{fig:exampleIA_d}) also violates the specification with robustness $-1$, but with output robustness $-2$. 
The classical robustness corresponds to the distance between the input $req$ and the threshold -- intuitively, the least expensive way to achieve the satisfaction of the specification is to decrease the amplitude of the input until the specification is vacuously satisfied. 
However, when measuring output robustness, the input is fixed and cannot change, thus yielding the value of $-2$, which corresponds to the distance between the output $gnt$ and the threshold. 
To enable output robustness and input vacuity, we first lift the notion of robustness $\rob$, to the more general notion of $U$-robustness relative to $V$, denoted by $\rob_U^V$, where $U\subseteq Y \subseteq X$ and $V\subseteq Y \subseteq X$, and $U \cap V = \emptyset$.
We define the robust semantics $\rob_U^V(\fun, \signal, t)$ by induction, where the only difference from the definition of $\rho(\fun, \signal, t)$ is the case of numeric predicates:
\begin{align*}
    &\rho_U^V(f(Y) > c, \signal, \timeElement) \notag\\
    &=\begin{cases}
	0 &\text{if } Y \not \subseteq U \cup V \\
	fw_Y(\timeElement))-c &\text{else if } Y \not \subseteq V \\
	\text{sign}(f(w_Y(\timeElement))-c) \cdot \infty &\text{otherwise}
    \end{cases}
\end{align*}

\noindent where for all $a \in \Reals$, $\text{sign}(a) \cdot \infty =+\infty$ if $a > 0$, $-\infty$ otherwise. Intuitively, $\rob(\fun, \signal ,\timeElement)$ measures the robustness of a specification to the signals in $U$ relative to the signals in $V$.
Hence, we use the classical robustness computation to compute predicates over variables in $U$, but we treat predicates over variables in $V$ and $Y\backslash(U \cup V)$ differently.
The predicates over variables in $V$ are given qualitative evaluation, resulting in robustness that can be either $+\infty$ or $-\infty$. The rationale is that we consider signals in $V$ to be \emph{fixed} and that a change resulting in switching the satisfaction or the violation of the predicates results in an infinite cost.
On the other hand, we consider signals in $Y\backslash(U \cup V)$ to be 
\emph{uncontrollable} and that even an infinitesimally small change could result in switching the satisfaction status of the predicate.

We are now ready to define the syntax and semantics of \iastl.

\begin{definition}[Interface-Aware Signal Temporal Logic]
An Interface-Aware Signal Temporal Logic (\iastl) specification over $X$ is a tuple $(X_U, X_V, \fun)$, where $X_U, X_V \subseteq X$, $X_U \cap X_V = \emptyset$ and $\fun$ is a \stl\ formula.
\end{definition}

We define \emph{output robustness} and \emph{input vacuity} using the notion of relative robustness:
\begin{description}
    \item[Output robustness] that is denoted $\mu$, as the $X_V$-robustness relative to $X\backslash X_V$.
    \item[Input vacuity] that is denoted $\nu$, the $X_U$-robustness relative to $\emptyset$.
\end{description}

\section{Architecture}
\label{sec:architecture}

\begin{figure}[t]
    \centering
	\includegraphics[width=0.9\linewidth]{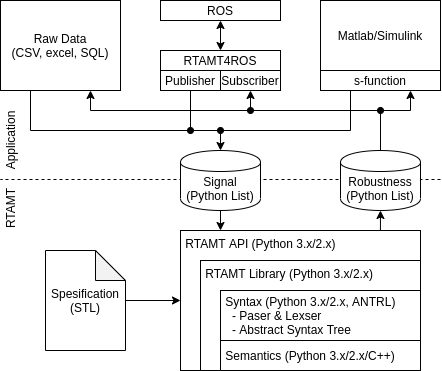}
    \caption{Overview of \rtamt.}
    \label{fig:overview}
\end{figure}

In this section, we introduce the overview of the \rtamt\ architecture for specification-based monitoring. Fig.~\ref{fig:overview} depicts the overview of \rtamt.
The core module of \rtamt\ API is developed in \python\footnote{Both \python\ 2.x and 3.x are supported for \ros\ and general purposes.}. for several reasons:
\begin{enumerate}
    \item Facilitates handling common input data formats such as CSV, Excel, or SQL server.
    \item Easy integration with \ros,  a well-known middleware in the robotics domain.
    \item Better connectivity with \matlabSimulink, a well-known Model Based Development (MBD) platform in the control design domain.
    \item Sophisticated software package system: Python Package Index (PyPI) enables smooth distribution.
    \item Popularity and thereby increasing user base.
\end{enumerate}

\begin{sidewaysfigure*}
    \centering
	\includegraphics[width=\linewidth]{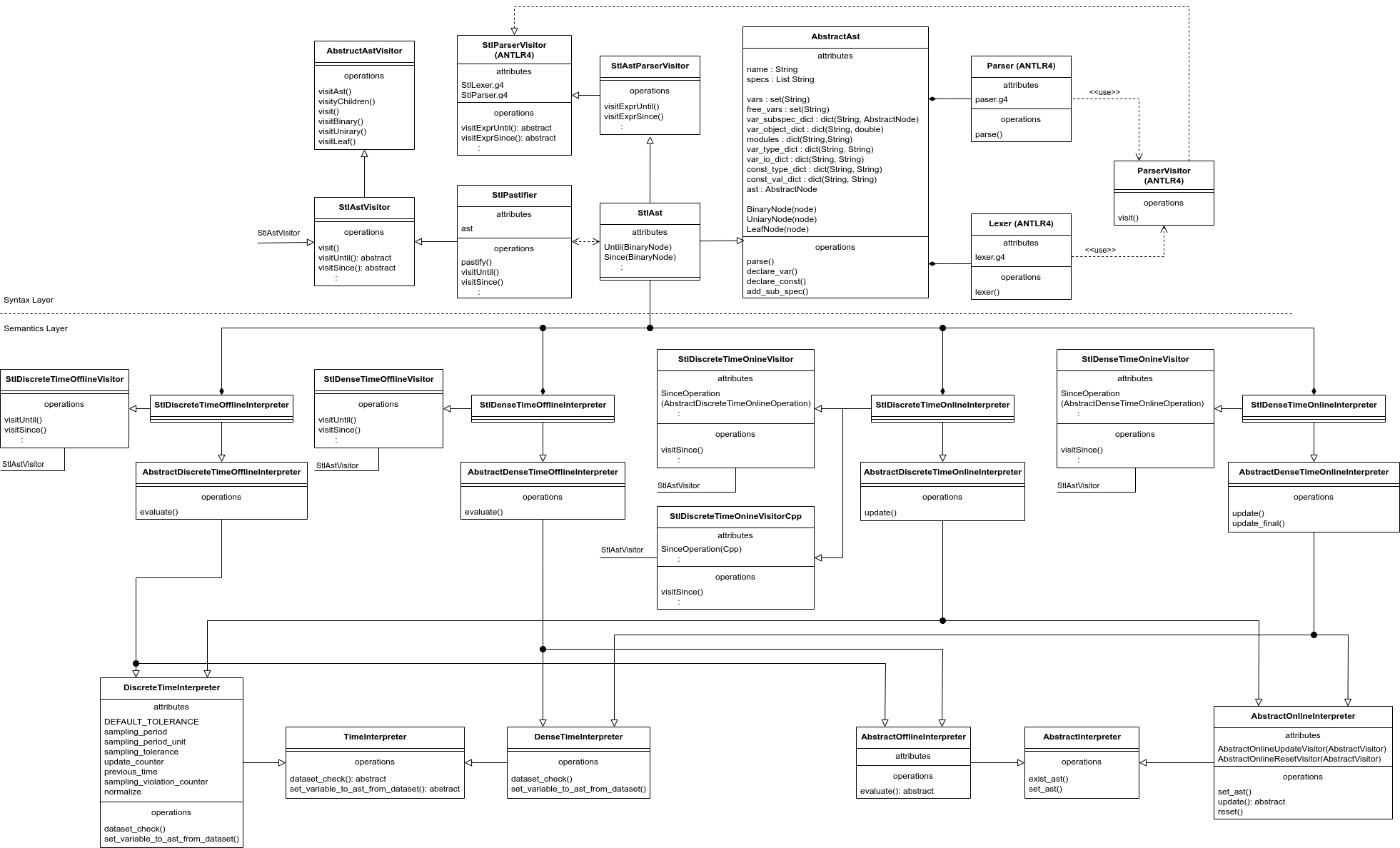}
    \caption{Architecture of \rtamt.}
    \label{fig:architecture}
\end{sidewaysfigure*}
For practitioners, we provide a flexible and modular library behind the API that enables them to implement their own specification language based on \ltl/\stl.
Fig.~\ref{fig:architecture} shows the architecture of the \rtamt\ Library in a class diagram.
The cores of the library are the syntax layer, which manages specification with lexer, parser, and Abstract Syntax Tree (\AST), and the semantic layer, which calculates robustness from given trajectories with specific semantics.
Since Python is an interpreted language, there is a natural trade-off between computing speed and ease of programming.
For this reason, we implement the syntax layer in \python, because the layers are executed only one time in the parse phase, and we prioritize flexibility rather than computing efficiency.
In contrast to the syntax layer, the semantic layer is called upon every monitoring update.
Because we expect real-time monitoring, we can exploit not only \python\ for rapid prototyping, but also \cpp\ for fast computing in this layer.
Next we explain each layer in detail.

\subsection{Syntax Layer}
We first utilize \antlr~\cite{parr2013definitive} lexer and parser to parse a given specification.
\antlr\ provides a flexible, easy-to-use platform for lexer and parser instantiated with a tool-specific setting file \antlr\ grammar.
{\tt StlAst} and {\tt StlAstVisitor} are instances of the \rtamt\ \AST\ and visitor class in \stl\ case as the core of this layer and the link to the next semantics layer.

\begin{figure*}[t]
    \centering
	\includegraphics[width=0.7\linewidth]{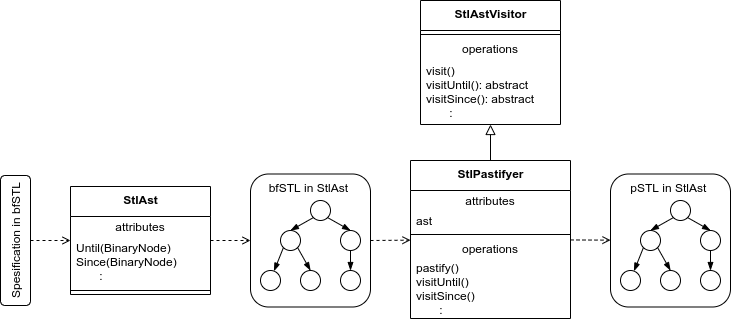}
    \caption{Pastification from \bfstl\ to \pstl.}
    \label{fig:stl-pastification}
\end{figure*}
Another functionality in this layer is handling any syntactic manipulations of the specification parse trees, including the pastification procedure from Sec.~\ref{sec:pastification}.
In that case, we translate the \bfstl\ formula $\phi$ into an equi-satisfiable \pstl\ formula $\psi$, which uses only past temporal operators.
The actual pastification parser {\tt StlPastifier} is extended from {\tt StlAstVisitor}, which provides the parse method for translating the specification text into an internal representation.
This pastification two-step process is depicted in Fig.~\ref{fig:stl-pastification}. 

\subsection{Semantics Layer}
This semantics layer provides algorithms for online and offline monitors from declarative specifications with quantitative semantics that are inductively implemented by traversing the \AST\ from the syntax layer.

The implementation of the monitoring algorithm is specific to the specification language used.
The time handler ({\tt TimeInterpreter}) handles two signal classes.
\begin{description}
    \item[Discrete-time {\rm({\tt DiscreteTimeInterpreter})}] This class is the base for monitoring discrete-time signals. The implementation follows a time-triggered approach in which sensing of inputs and output generation are done at a periodic rate. Its use is motivated by~\cite{henzinger1992good}, which shows that by weakening/strengthening real-time specifications, discrete-time evaluation of properties preserves important properties of dense-time interpretation.
    This approach admits an upper bound on the use of computation resources.
    The discrete-time monitors essentially implement the algorithm from~\cite{ptstl} adapted to the robustness semantics.
    
    \item[Dense-time {\rm({\tt DenseTimeInterpreter})}] This class is the base for monitoring dense-time signals, which use piece-wise constant interpolation.
    The dense-time monitors implement the algorithms from~\cite{robust2} and adapt them to the piece-wise constant interpretation of signals. 
    The resulting monitors follow an event-driven approach, where the samples can happen at any time in an intervals of reals and there can be an arbitrary (but finite) number of samples in any unit interval of time.
    This approach is suitable for more advanced and distributed applications in which state variable observations are not periodically triggered but rather driven by external, hence uncontrollable events. The key algorithmic ingredient in the offline evaluation of STL interpreted over dense-time signals is an optimal streaming algorithm for computing the minimum (or the maximum) values of numeric sequences (samples) over a sliding window of fixed size.
    This procedure allows to compute efficiently temporal operators $\globally_I$, $\finally_I$, $\historically_I$ and $\once_I$, whose semantics is defined as a computation of the minimum/maximum values over a sliding window $I$. It is also adapted to enable the evaluation of the more general $\until_I$ and $\since_I$ operators.
    To address the online monitoring problem, we use the incremental application of the offline evaluation approach from~\cite{amt} to the partially received inputs.
\end{description}

In addition to the above, we support two monitoring modes.
\begin{description}
    \item[Offline Monitor {\rm({\tt AbstractOfflineInterpreter})}] This class supports typical offline evaluation, which expects all data to exist at once when evaluated.
    This provides a method {\tt evaluate} to evaluate the signal.

    \item[Online Monitor {\rm({\tt AbstractOnlineInterpreter})}] This class supports online evaluation of signals with different timings.
    The operation thus needs to manage memory to evacuate not incomplete signals until the next signal update.
    This provides a method {\tt update} to evaluate signals.
\end{description}
Both {\tt evaluate} and {\tt update} method take a list of variable names, {\tt (time, value)} pairs and return a float number representing the robustness degree of the formula at that time index relative to the input prefix observed in the present.


This layer supports a \cpp\ implementation to enable real-time as well.
We use the Boost \python\ library, a \cpp\ library which enables seamless operability between the two languages, to integrate the \python\ front-end with the \cpp\ back-end.


\subsection{Integration of RTAMT to \ros}
\begin{figure*}[t]
    \centering
	\includegraphics[width=0.6\linewidth]{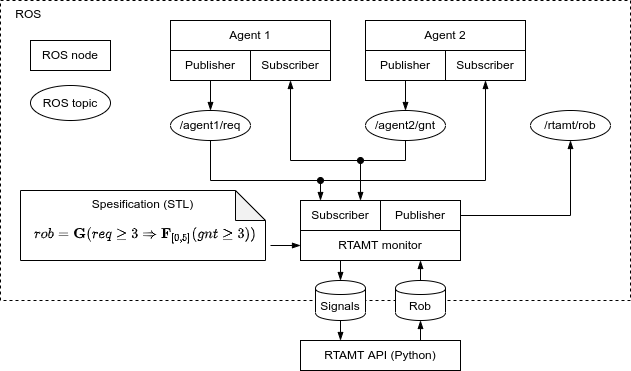}
    \caption{Integration of \rtamt\ to \ros.}
    \label{fig:ros}
\end{figure*}
\ros\ is the de-facto standard in developing robotic applications, 
supporting  several messaging approaches. 
In this paper, we assume that \ros\ nodes use the subscriber/publisher messaging pattern.
A publisher categorizes a message into a class (called a topic in \ros) and sends it without any knowledge of who will read the message. 
A subscriber expresses interest in receiving message from one or more topics and only receives messages of interest, without knowing who sent the message\footnote{Unless the publisher encodes its identity into the message itself.}. 
We also assume that messages are published at a periodic (and known) rate.

The integration of \rtamt\ into \ros\ is done in {\tt rospy}, as illustrated in Fig.~\ref{fig:ros}.
We assume that we are monitoring a \ros\ system consisting of one or more nodes that publish messages via variables appearing in the \bfstl\ specification. 
The monitoring \ros\ node uses \rtamt\ to parse and evaluate the \bfstl\ formula.
In order to communicate with other \ros\ nodes that publish the relevant messages, the specification must associate variable names with their associated \ros\ topic names.
This is done using annotations.
The specification defines which topics the monitor needs to subscribe and publish (topics {\tt rtamt/req}, {\tt rtamt/gnt}, and {\tt rtamt/out} in our example).
Since the names of the variables and associated topics as well as the variable types are not known in advance but are inferred from the specification definition, we implemented a dynamic subscribing and publishing mechanism using \python\ introspection and reflection. 
For each variable that appears in the specification, we check whether it has a valid \ros\ message type, generate an instance of the object and retrieve its class information.
This information provides a dynamic subscriber and publisher.
In addition, a \rtamt\ monitor is called by a single dynamic callback function as a \ros\ node and publishes the robustness with the new incoming data from the subscriber.

An existing example of \stl\ evaluation in \ros\ is shown in Sec.~\ref{sec:evaluating_ROS}, and an experiment is shown in Sec.~\ref{sec:experiment_ROS}.

\subsection{Integration of RTAMT to \matlabSimulink}
\label{sec:intsim}
We integrate \rtamt\ into the \matlabSimulink\ environment via S-functions (short for {\textsf system functions}). 
S-functions provide a powerful mechanism to extend the existing capabilities of the Simulink environment by enabling to program custom Simulink blocks.
S-functions use a special API that allows interaction with the simulation engine similar to how the built-in Simulink blocks interact with the engine. 
More specifically, we use MATLAB Level 2 S-functions that allow us to integrate the \rtamt\ API using the MATLAB m-scripts.

An S-function consists of a set of callback methods that are customized to provide the desired functionality. 
The Simulink engine calls the appropriate method at each stage of the simulation.
The main S-function callback methods are associated with the following tasks:
\begin{description}
    \item[Compilation] The stage in which the \matlabSimulink\ engine initializes the S-function and its parameters.
    \item[Calculation of outputs] at this state, the engine computes the outputs of the block until all its output ports are valid for the current time step.
    \item[Update discrete states] The block performs activities that are executed only once per time step such as updating discrete states.
    \item[Initialize and Terminate Methods] Activities required by S-function only once that are performed at the beginning (setting up practitioner's data, initializing state vectors, etc.) and at the end (memory deallocation, etc.) of the simulation. 
    \item[Integration] Computation related to the continuous states and/or nonsampled zero crossings at minor time steps. 
\end{description}

The integration of \matlabSimulink\ consists of the following steps:
\begin{itemize}
    \item We define the \rtamt\ online monitor S-function block with a single input port consisting of an array of double signals (one for each variable appearing in the specification), a single output signal (the robustness signal) and a single block parameter (the \iastl\ specification).
    \item In the initialize method of the S-function, we create the \rtamt\ \stl\ specification object, parse the specification passed as a parameter to the block and pastify the monitor.
    \item In the method for calculating output, we fetch the most recent array of input data from the input port of the post, pass it to the monitor update function, and forward it to the robustness output port.
\end{itemize}

An existing example of \stl\ evaluation on \matlabSimulink\ is shown in Sec.~\ref{sec:evaluating_Simulink}, and an experiment is shown in Sec.~\ref{sec:experiment_Simulink}.

\section{Application Programming Interface}
\label{sec:api}

In this section we show a basic-use case of the \rtamt\ API in \python\ with time-stamped data and evaluate it with an arbitrary specification.
The three main steps in defining a specification are as follows:
\begin{enumerate}
    \item define: method for accessing a practitioner-defined specification from a file,
    \item parse: method for parsing the specification and building its internal representation,
    \item evaluate: method for passing the next snapshot of the input variables and computing the resulting robustness.
\end{enumerate}

\subsection{Evaluating \stl\ in Offline}
\begin{lstlisting}[language=Python, caption=Evaluating \stl\ in offline., label=code:offlineSTL]
import rtamt

spec = rtamt.StlDenseTimeOfflineSpecification()
spec.declare_var('req', 'float')
spec.declare_var('gnt', 'float')
spec.spec = 'G((req>=3)->(F[0,5](gnt>=3)))'
spec.parse()

req = [[0.0, 0.0], [2.0, 6.0], [4.0, 0.0], [10.0, 0.0]]
gnt = [[0.0, 0.0], [6.0, 6.0], [8.0, 0.0], [10.0, 0.0]]

rob = spec.evaluate(['req', req], ['gnt', gnt])
\end{lstlisting}
Listing~\ref{code:offlineSTL} illustrates an example code of an offline monitor.
It follows the above steps. 

\vspace{10pt}
\subsection{Evaluating \stl\ in Online}
\begin{lstlisting}[
    language=Python,
    caption=Evaluating \stl\ in online.,
    label=code:onlineSTL
]
import rtamt

spec = rtamt.StlDenseTimeOnlineSpecification()
spec.declare_var('req', 'float')
spec.declare_var('gnt', 'float')
spec.spec = 'H[0,10]((O[5,5](req>=3))->(O[0,5](gnt>=3)))'
spec.parse()

req_0 = [[0.0, 0.0], [2.0, 6.0]]
gnt_0 = [[0.0, 0.0], [5.0, 6.0]]
req_1 = [[4.0, 0.0], [10.0, 0.0]]
gnt_1 = [[8.0, 0.0], [10.0, 0.0]]

rob = spec.update(['req', req_0], ['gnt', gnt_0])
rob = spec.update(['req', req_1], ['gnt', gnt_1])
\end{lstlisting}
Listing~\ref{code:onlineSTL} illustrates an example of online monitor code.
Major differences from offline use are that only \pstl\ is allowed for the specification and the evaluation is done by periodic {\tt update()}.
We note that pastification can convert \bfstl\ to \pstl\ if we insert {\tt spec.pastify()} after line 7 {\tt spec.parse()}.

\vspace{10pt}
\subsection{Evaluating \stl\ on \ros}
\label{sec:evaluating_ROS}
We sketch the \ros\ interface to \rtamt in \rosrtamt.

\begin{lstlisting}[
    language=Python,
    caption=Skeleton of the \rosrtamt\ interface., label=code:ROS,
    linerange={72-90}
]
def monitor(period, unit):
    # Init STL SPEC
    spec = init_spec(period, unit)
    # Init ROS node
    rospy.init_node(spec.name, anonymous=True)
    # Init subscriber and publisher
    pub = sub_and_pub(spec)
    # Set monitoring frequency
    rate = rospy.Rate(spec.get_sampling_frequency())
    # Control loop
    time_index = 0
    while not rospy.is_shutdown():
        rob_msg = mon_update(spec, time_index)
        pub.publish(rob_msg)
        time_index += 1

        # Wait until next evaluation
        rate.sleep()
\end{lstlisting}
The skeleton of the interface, defined by the \emph{monitor} procedure, is shown in Listing~\ref{code:ROS}. 
The first step consists in initializing the \stl specification that is to be monitored (line 3).
The second step consists in initializing the \ros\ node that hosts the \stl\ monitor (line 5), subscriber, and publisher (line 7), then sets the monitoring frequency (line 9).
The \stl\ monitoring node than enters the infinite monitoring loop, which is periodically invoked at the specified frequency (line 18). 
Each iteration of the loop consists of three steps:
\begin{enumerate}
    \item automatic invocation of the callback mechanism (Listing~\ref{code:callback}),
    \item invoking the monitoring update (line 13, Listing~\ref{code:update}), and 
    \item publishing the robustness to the appropriate topic (line 14).
\end{enumerate}

\begin{lstlisting}[
    language=Python,
    caption=Callback procedure.,
    label=code:callback,
    linerange={32-37}
]
def callback(data, args):
    spec = args[0]
    var_name = args[1]

    var = copy.deepcopy(data)
    spec.var_object_dict[var_name] = var
\end{lstlisting}
The callback procedure, shown in Listing~\ref{code:callback}, simply receives a new input sample message and copies it to the monitor.

\begin{lstlisting}[
    language=Python,
    caption=Specification initialization procedure.,
    label=code:init,
    linerange={9-29}
]
def init_spec(period, unit):
    spec = rtamt.StlDiscreteTimeOnlineSpecificationCpp()
    spec.set_sampling_period(period, unit)

    spec.name = 'HandMadeMonitor'
    spec.import_module('rtamt_msgs.msg', 'FloatStamped')
    spec.declare_var('req', 'FloatStamped')
    spec.declare_var('gnt', 'FloatStamped')
    spec.declare_var('rob', 'FloatStamped')
    spec.set_var_topic('req', 'rtamt/req')
    spec.set_var_topic('gnt', 'rtamt/gnt')
    spec.set_var_topic('rob', 'rtamt/gnt')
    spec.spec = \
        'rob.value = G[0,10]((req.value>=3)->(F[0,5](gnt.value>=3)))'

    try:
        spec.parse()
        spec.pastify()
    except rtamt.STLParseException as err:
        sys.exit()
    return spec
\end{lstlisting}
In the initialization procedure of specification (see Listing~\ref{code:init}), the practitioner creates an \stl\ specification monitor (line 2), sets the sampling period of the monitors (line 3), imports the \ros\ message type that will be processed by the monitor (line 6), defines the specification variables and the property to monitor (lines 7-12), and parses the specification (line 17).
The pastification is applied because the property is \bfstl (line 18).

\begin{lstlisting}[
    language=Python,
    caption=Subscription and publication to \ros\ topics.,
    label=code:subpub,
    linerange={40-53}
]
def sub_and_pub(spec):
    # Publish STL output var (robustness)
    topic = spec.var_topic_dict[spec.out_var]
    out = spec.get_value(spec.out_var)
    pub = rospy.Publisher(topic, out.__class__, queue_size=10)

    # Subscribe STL input vars
    for var_name in spec.free_vars:
        var_object = spec.get_value(var_name)
        topic = spec.var_topic_dict[var_name]
        rospy.Subscriber(topic, var_object.__class__, callback, [spec, var_name])

    return pub
\end{lstlisting}
The procedure shown in Listing~\ref{code:subpub} registers the output variable of the monitor to a publisher, and all the variables appearing in the \stl\ specification to a subscriber. 
We use reflection and introspection to dynamically determine the variables that need to be registered, and their types.

\begin{lstlisting}[
    language=Python,
    caption=Monitoring update.,
    label=code:update,
    linerange={55-71}
]
def mon_update(spec, time_index):
    var_name_object_list = []
    for var_name in spec.free_vars:
        var_name_object = (var_name, spec.get_value(var_name))
        var_name_object_list.append(var_name_object)

    # Update the monitor
    # spec.update is of the form
    # spec.update(time_index,[('a',aObj),...)])
    rob_msg = spec.update(time_index, var_name_object_list)

    rob_msg.header.seq = time_index
    rob_msg.header.stamp = rospy.Time.now()

    return rob_msg

\end{lstlisting}
Finally, the monitor updates its evaluation and computes robustness, as shown in Listing~\ref{code:update}.

\subsection{Textual Specifications for \ros}
\label{sec:textual_example}
\begin{lstlisting}[
    language=Python,
    caption=The textual form "spec.stl".,
    label=code:spec1
]
// name
specification spec_x

// imports
from rtamt_msgs.msg import FloatStamped

// variable declarations
input FloatStamped req
output FloatStamped gnt
output FloatStamped rob

// annotations
@ topic(req, rtamt/req)
@ topic(gnt, rtamt/gnt)
@ topic(rob, rtamt/rob)

// bfSTL property
rob.value = G[0,10]((req.value>=3)->(F[0,5](gnt.value>=3)))
\end{lstlisting}
The tool supports an automatic monitor setting for \ros\ with a textual form for non-programmer practitioners.
The concrete textual specifications consist of: the name header, a list of \python\ modules to be imported, a list of variable declarations, a list of annotations, and the property.
Listing~\ref{code:spec1} formulates this \stl\ property.
The syntax of the specification language allows the use of arbitrary data types.
We assume that variables {\tt req} and {\tt gnt} are defined as objects of \python\ type {\tt FloatStamped} with attributes {\tt header} of type {\tt Header} and {\tt value} of type {\tt float}
\footnote{
    We recall that \bfstl\ is defined over real-valued variables. 
    As a consequence, we allow \bfstl\ properties to refer to \python\ variables and attributes of type {\tt int}, {\tt long} and {\tt float}
}.
We also need to import the module that defines the arbitrary data type used in the specification (see line 5).
The annotations are special comments that are by default ignored but that can provide additional information in specific contexts.
For now we ignore the annotations from lines 13-15. 
We finally note that the specification\footnote{
    We allow only \pstl\ and \bfstl. The pastification is applied in the background automatically when practitioners use \bfstl.}
robustness is assigned to one of the declared variables ({\tt rob.value} in line 18).
Practitioners can launch these textual specifications in \ros\ below,
\begin{lstlisting}[language=bash]
$ rosrun rtamt4ros ros_stl_monitor.py spec.stl --period 1 --unit s
\end{lstlisting}

\subsection{Evaluating \stl\ in \matlabSimulink}
\label{sec:evaluating_Simulink}
\begin{figure*}
    \centering
    \includegraphics[width=0.8\linewidth]{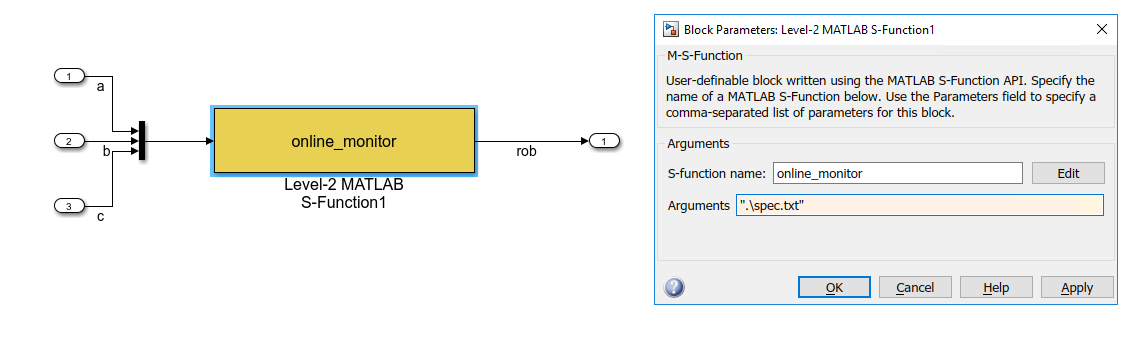}
    \caption{Integration of \rtamt\ in \matlabSimulink.}
    \label{fig:apisim}
\end{figure*}
Now we describe how \rtamt\ monitors can be used in \matlabSimulink\ models and illustrate the API in Fig.~\ref{fig:apisim}. 
The integration is done by inserting the {\tt online{\_}monitor} block into an existing Simulink model.
The monitor is a MATLAB Level 2 S-function block that integrates \rtamt\ as explained in Sec.~\ref{sec:intsim}. 
The block has one input and one output port. 
In order to pass multiple input signals to the monitor via the single input port, all the input signals must be combined into a single array signal using the {\tt multiplexer} block.
The output of the block is the robustness signal, which can be used to visually inspect the robustness of the property but also for any other appropriate purpose.

\section{Implementations of \stl\ Extended Language}
\label{sec:library}
In this section, we describe how practitioners implement a new specification with the \rtamt\ library.
We first sketch how to extend STL monitors with the IA-STL syntax and semantics. The aim is to reuse as much as possible of the RTAMT STL monitoring implementation and only add implementation where STL and IA-STL differ. We first identify and localize main differences between STL and IA-STL:
\begin{itemize}
\item Syntax: IA-STL extends STL with the ability to declare signal variables as \emph{input} and \emph{output}.
\item Semantics: the only difference in the semantics of IA-STL compared to STL is the treatment of numerical predicates.
\end{itemize}

The syntax extension requires adding the \emph{input} and \emph{output} keywords to the lexer and adding an appropriate parsing rule for allowing to decorate variables with these keywords.
To add the semantic extension, the practitioner must extend the visitor used to traverse the STL formula parse tree during its evaluation and overwrite the method that evaluates the numerical predicates.

The second scenario is the addition of a new operator to \stl.
In this case, practitioners also need to change both syntax and semantics. First, the name of the operator must be added as a reserved word to the lexer, as well as a new parsing rule that describes its signature. In contrast to the previous scenario, the visitor that parses the formula during evaluation must be extended with a new rule that provides the evaluation algorithm for the new operator.
Since the \rtamt\ library provides support functionalities in abstract classes, practitioners can implement a new specification monitor quickly and easily. 
\python\ enables fast prototyping whereas \cpp\ offers fast computation; the practitioner may choose according to their need.

\section{Experiments}
\label{sec:experiments}
We make experiments for \rtamt\ from various perspectives and in different environments: \rtamt\ unit testing, \ros, and \matlabSimulink.
All experiments are run on Intel\textsuperscript{\textregistered}~i9-10900K, 3.7~[GHz], 10 cores, and 128~[GB] RAM on Ubuntu 18.04.

\subsection{Comparison of Computational Efficiency}
In this section, we empirically evaluate the computational efficiency of the \rtamt\ in different settings. 

\begin{figure*}[t]
   \begin{minipage}{0.46\hsize}
        \centering
        \includegraphics[width=\linewidth]{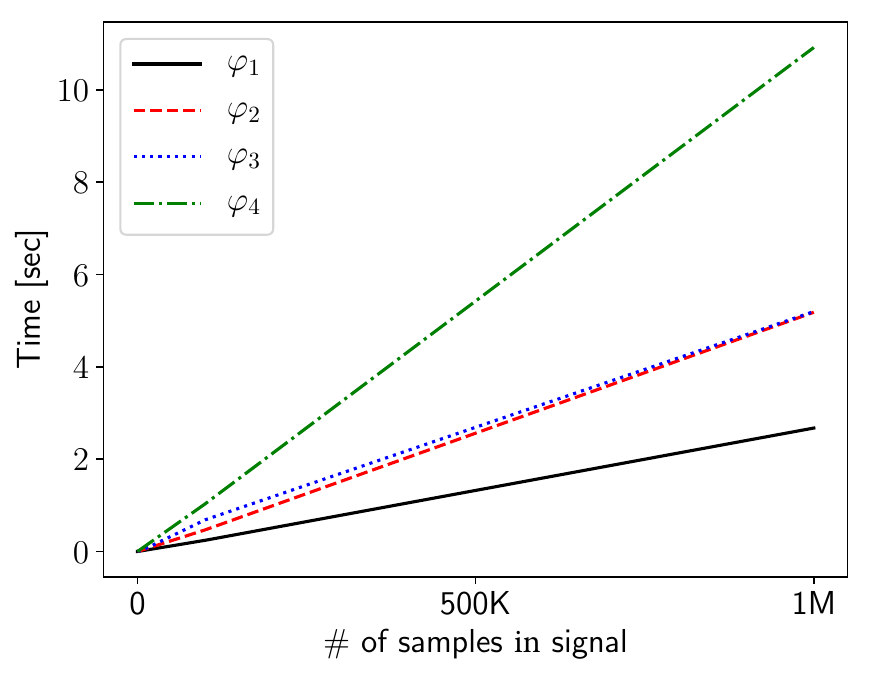}
        \subcaption{Discrete-time}
        \label{fig:formula-input-scaling_discrete}
    \end{minipage}
    \hspace{24pt}
    \begin{minipage}{0.46\hsize}
        \centering
        \includegraphics[width=\linewidth]{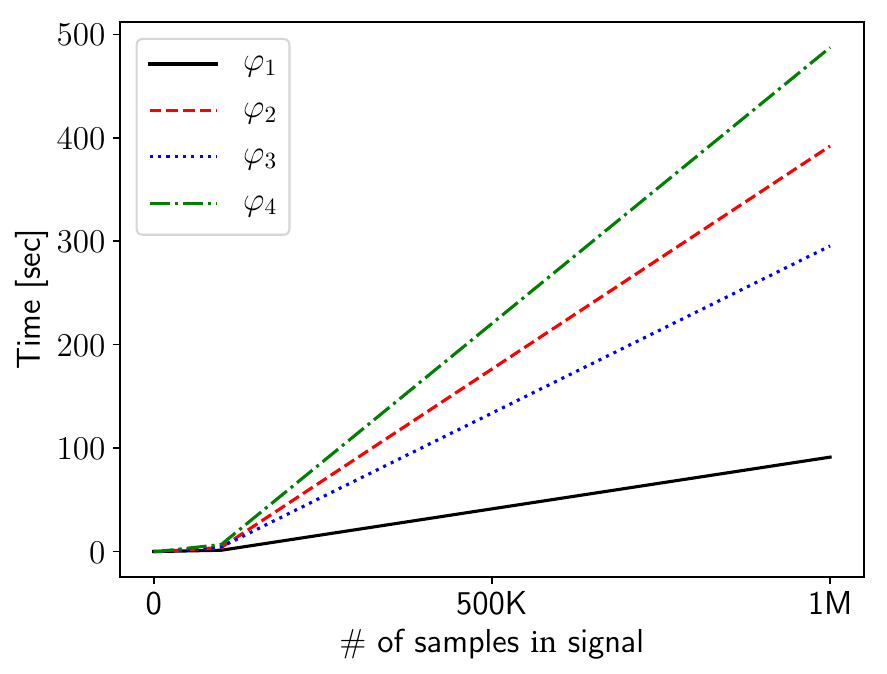}
        \subcaption{Dense-time}
        \label{fig:formula-input-scaling_dense}
    \end{minipage}
    \caption{Scaling to the number of samples in the input signal and the length of formulas in the offline monitors:~
    $\varphi_1 \equiv req\geq3$,~
    $\varphi_2 \equiv req\geq3 \imply gnt\geq3$,~
    $\varphi_3 \equiv req\geq3 \imply \finally_{[0,5]}gnt\geq3$,~
    $\varphi_4 \equiv \historically((req\geq3) \imply \neg (req\geq3)\until_{[0,5]}(gnt\geq3))$. 
    Each experiment was repeated 50 times, and we report the average computation time.}
    \label{fig:formula-input-scaling}
\end{figure*}
We first show how \rtamt\ scales in the size of the input trace and the length of formulas in the offline monitors of the \python\ back-end.
The results are depicted in Fig.~\ref{fig:formula-input-scaling} and show the difference of the average calculation time between discrete-time and dense-time monitors.

\begin{figure}[t]
    \centering
	\includegraphics[width=0.95\linewidth]{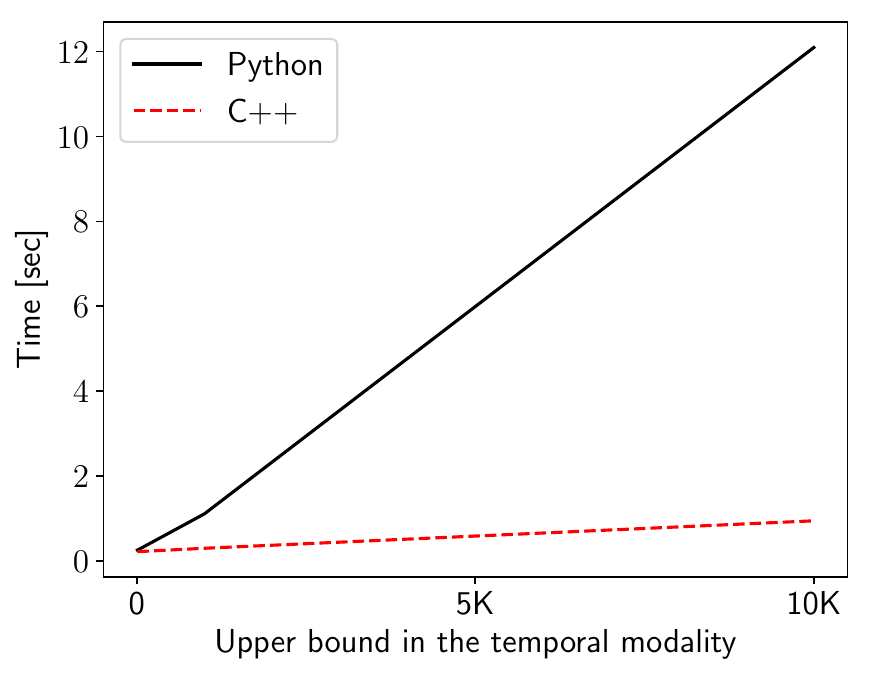}
    \caption{Comparison between \python\ and \cpp\ in the online monitors.}
    \label{fig:pythonCpp}
\end{figure}
Fig.~\ref{fig:pythonCpp} compares the computation time of the online monitors of the \cpp\ back-end against the \python\ back-end.
To compare the two algorithms, we use for the experiment the \stl\ specification $\globally_{[0,k]} (a+b \geq -2)$ where $k$ is the upper bound on the timing modality of the always operator, which we vary between 100 and 1~million.
The outcomes clearly demonstrate more than approximately 10 times better efficiency of the \cpp\ back-end, especially for large upper bounds in temporal modalities.

Those experiences guide the choice of the monitors for practitioners.
However, even the slowest case which is dense-time of the \python\ back-end calculates approximately 0.5~[ms] per a sample and has a good enough real-time performance.

\subsection{\hsr\ Simulator in \ros}
\label{sec:experiment_ROS}

\begin{figure*}[t]
    \centering
    \begin{minipage}{0.32\hsize}
        \centering
        \includegraphics[width=\linewidth]{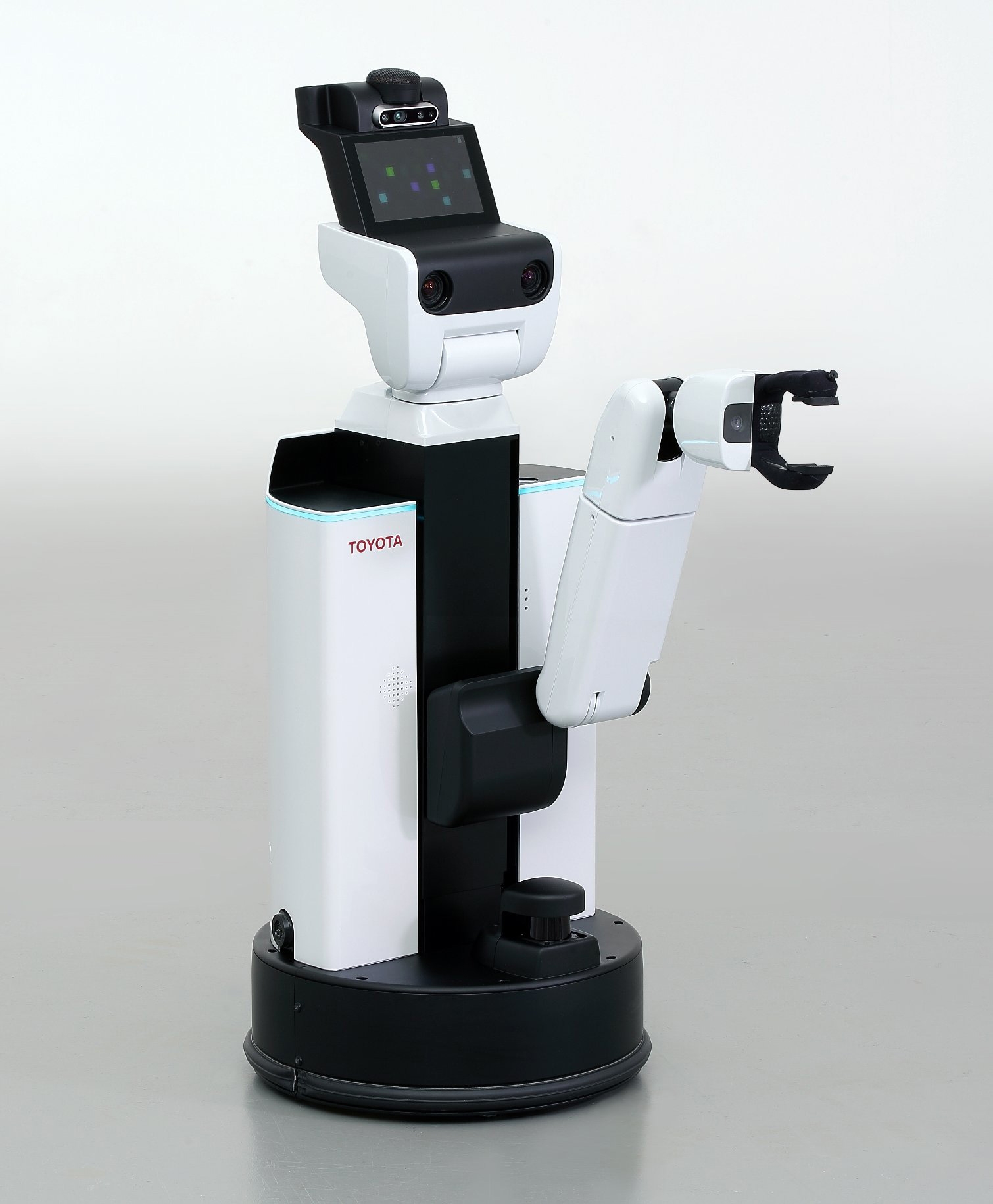}
        \caption{\hsr.}
        \label{fig:HSR}
    \end{minipage}
    \begin{minipage}{0.67\hsize}
        \centering
        \captionof{table}{\hsr\ basic specification.}
        \label{table:HSRspec}
        \scriptsize
        \begin{tabular}{|l|l|l|}
            \hline
            Size & Footprint & $\phi$450 [mm]\\
                & Height (min/max) & 1,047/1,392 [mm] \\
                & Freedom & 8 \\
                & Weight & About 40 [kg] \\ \hline
            Base & Drive system & Omnidirectional \\
                & Environment & Indoor step 10 [mm] \\
                &             & Climbing 7 [deg] \\ 
                & Max speed   & 0.22 [m/s] \\ \hline
            Hoisting & Stroke & 690 [mm] \\
                & Max speed & 150 [mm/s] \\ \hline
            Arm & Length & About 600 [mm] \\
                & Movable height & 0\textasciitilde1,350 [mm] \\
                & Movable depth & 450 [mm] \\
                & Payload & 0.5 [kg] \\ \hline
            Gripper & Grip speed & $<$0.4 [s] \\
                & Max force & About 40 [N] \\
                & Grip range & 135 [mm] \\
                & Max suction force & About 4 [N] \\ \hline
        \end{tabular}
    \end{minipage}
\end{figure*}

We now show that \rtamt\ not only monitors safety specification,  but also enables fault localization on the component level in an autonomous mobile system in an \ros\ environment. 
This experiment follows authors' previous work~\cite{yamaguchi2020specification} which is based on assume-guarantee fashion.

We applied \rosrtamt\ to Human Support Robot (\hsr)~\cite{yamamoto2019development} (Fig.~\ref{fig:HSR}), provided by Toyota as a robot platform. 
Physically, \hsr\ has 8~Degrees of Freedom (DoF), combining the 3~DoF of the mobile base, 4~DoF of the arm and 1 DoF of the torso lift~(Table~\ref{table:HSRspec}), as well as several (Light Detection And Ranging) LiDARs, along with Stereo and monocular cameras.
The platform supports \ros~Gazebo~\cite{Koenig-2004-394} simulator.
We applied \rosrtamt\ to the simulation.

\begin{figure*}[t]
    \centering
    \begin{minipage}{0.50\hsize}
        \centering
        \includegraphics[width=\linewidth]{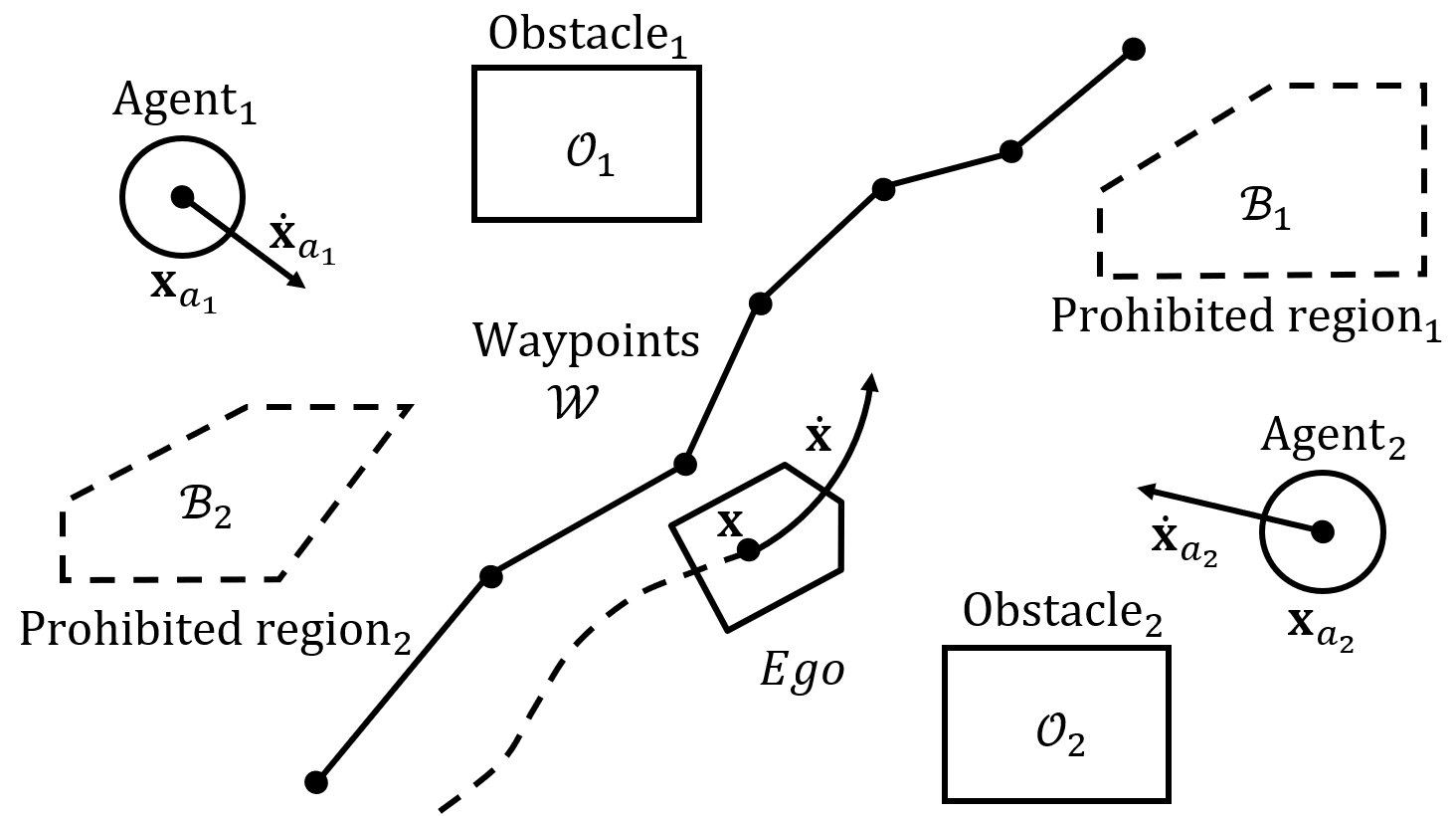}
        \subcaption{Variable map.}
        \label{fig:variableMap}
    \end{minipage}
    \hspace{12pt}
    \begin{minipage}{0.39\hsize}
        \centering
        \includegraphics[width=\linewidth]{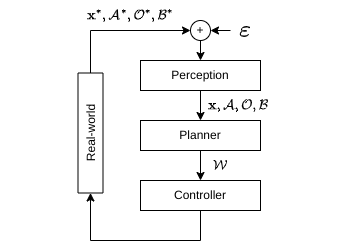}
        \subcaption{Architecture.}
        \label{fig:targetArchtecture}
    \end{minipage}

    \caption{\hsr\ system: (a) ego position $\pos$, agents' position $\agents = \{\agentpos{1},\ldots,\agentpos{\mid \agents \mid}\}$, static obstacles' position $\obstacles = \{\obstacle{1},\ldots,\obstacle{\mid \obstacles \mid}\}$, and prohibited regions $\prohibregions = \{\prohibregion{1}\ldots,\prohibregion{\mid \prohibregions \mid}\}$
    (b) ground truth of $\ground{\pos}$, $\ground{\agents}$, $\ground{\obstacles}$, and $\ground{\prohibregions}$ from real-world plant with error $\varepsilon$.}
\end{figure*}

Fig.~\ref{fig:targetArchtecture} provides an overview of the target's software architecture.
The simple perception, planner, and controller layers constitute the basic and abstract architecture in an autonomous mobile system \cite{urmson2008autonomous} based on the \ros\ software platform.
Here is a brief description of each subsystem.
\begin{description}
    \item[Perception] 
    In an actual robot, such as a range sensor, radar and vision-based perception are commonly used with sensor fusion techniques such as Kalman-filters or Particle-filters for detection and localization~\cite{thrun2005probabilistic}.
    Specifically, the \hsr\ perception layer has 2D grid map-based localization from the \ros\ navigation package~\cite{tbd}.
    In formalism, the perception recognizes statuses: ego position $\pos=(x,y,\theta)$ where 2D position $x$, $y$, and angular position $\theta$, agents' position $\agents = \{\agentpos{1},\ldots,\agentpos{\mid \agents \mid}\}$, static obstacles' position $\obstacles = \{\obstacle{1},\ldots,\obstacle{\mid \obstacles \mid}\}$, and prohibited regions $\prohibregions = \{\prohibregion{1}\ldots,\prohibregion{\mid \prohibregions \mid}\}$
    from ground truth of $\ground{\pos}$, $\ground{\agents}$, $\ground{\obstacles}$, and $\ground{\prohibregions}$ from real-world plant with error $\varepsilon$.
    
    \item[Planner] 
    This layer decides its own path $\waypoints=\{\waypoint{1},\ldots,\waypoint{\mid \waypoints \mid}\}$ based on the previous perception layer's recognition $\pos$, $\agents$, $\obstacles$, and $\prohibregions$.
    Generally, the robot has a global path planner and local path planner.
    The former maps an overview of a path from a high-level graph similar to a static car-navigation map; the latter generates a motion path that treats more closed and dynamic obstacles.
    In the \hsr\ planner layer, for simplicity, we omit the global path planner and deploy only Rapidly-exploring Random Trees (RRTs) \cite{kuffner2000rrt}  as a local path planner to reach a goal.
    
    \item[Controller]
    Typically, this layer handles low-level hardware such as vehicle base control and actuator control based on $\waypoints$ from the previous planner subsystem.
    Because it is a low-level control, generally a PID controller yields adequate results.
    In the case of \hsr,\, a PID-based path follower is implemented which tracks way-points, minimizing error between current position $\pos$ and $\waypoints$ from the planner.
\end{description}

Based on this architecture, we define safety specifications in the system and subsystems.
The most important system-level safety specification is a collision prohibition.
\begin{equation}
    \varphi_{sys} \equiv \bigwedge_{T_i \in \agents, \obstacles, \prohibregions}
    \globally_{[0,\tau]}\left(\Dist(\ground{\pos},\ground{T_i})>C_{coll}\right)
    \label{eq:systemSpec}
\end{equation}
Where $\Dist$ is the Euclidean distance and $\tau$ is the time window of the formula. 
Simply $\globally$ expresses "all time" the distance between the ego and others should be greater than some threshold. 
We use ground truths $\ground{\pos}$, $\ground{\agents}$, $\ground{\obstacles}$, and $\ground{\prohibregions}$ while take an advantage of using a simulator since $\pos$, $\agents$, $\obstacles$, and $\prohibregions$ have recognition errors of those.

The role of the perception subsystem is to recognize other obstacles and agents as much as possible.
In other words, we may check the discrepancy between any recognition and the ground truth as perception specifications:
\begin{align}
    &\varphi_{per} \equiv \notag\\
    &\bigwedge_{T_i \in \pos, \agents, \obstacles, \prohibregions}
    \globally_{[0,\tau_1]}\finally_{[0,\tau_2]}\left(\errfn(T_i,\ground{T_i}) < \varepsilon_{per} \right)
    \label{eq:perceptionSpec}
\end{align}
Where $\errfn$ is a perception error metrics in each specific domain.
We use $\globally\finally$ to give a tolerance of the violation with the time window $\tau_1$. $\tau_2$ is the duration time which permits the error since a small perception error is expected and it does not affect the system issue immediately.

The planner subsystem should generate a reasonable path $\waypoints$ which does not cause collision with others:  
\begin{equation}
    \varphi_{plan} \equiv \bigwedge_{T_i \in \agents, \obstacles, \prohibregions}
    \globally_{[0,\tau]}\left(\Dist(\waypoints,T_i)>C_{plan}\right)
    \label{eq:plannerSpec}
\end{equation}

Finally, the controller subsystem should follow the reference as a low level controller:
\begin{equation}
    \varphi_{con} \equiv \globally_{[0,\tau_1]}\finally_{[0,\tau_2]}\left(\errfn(\waypoints,\pos) < \varepsilon_{con}\right)
    \label{eq:controllerSpec}
\end{equation}

\begin{figure*}[t]
    \centering
    \begin{minipage}{0.47\hsize}
        \centering
        \includegraphics[width=\linewidth]{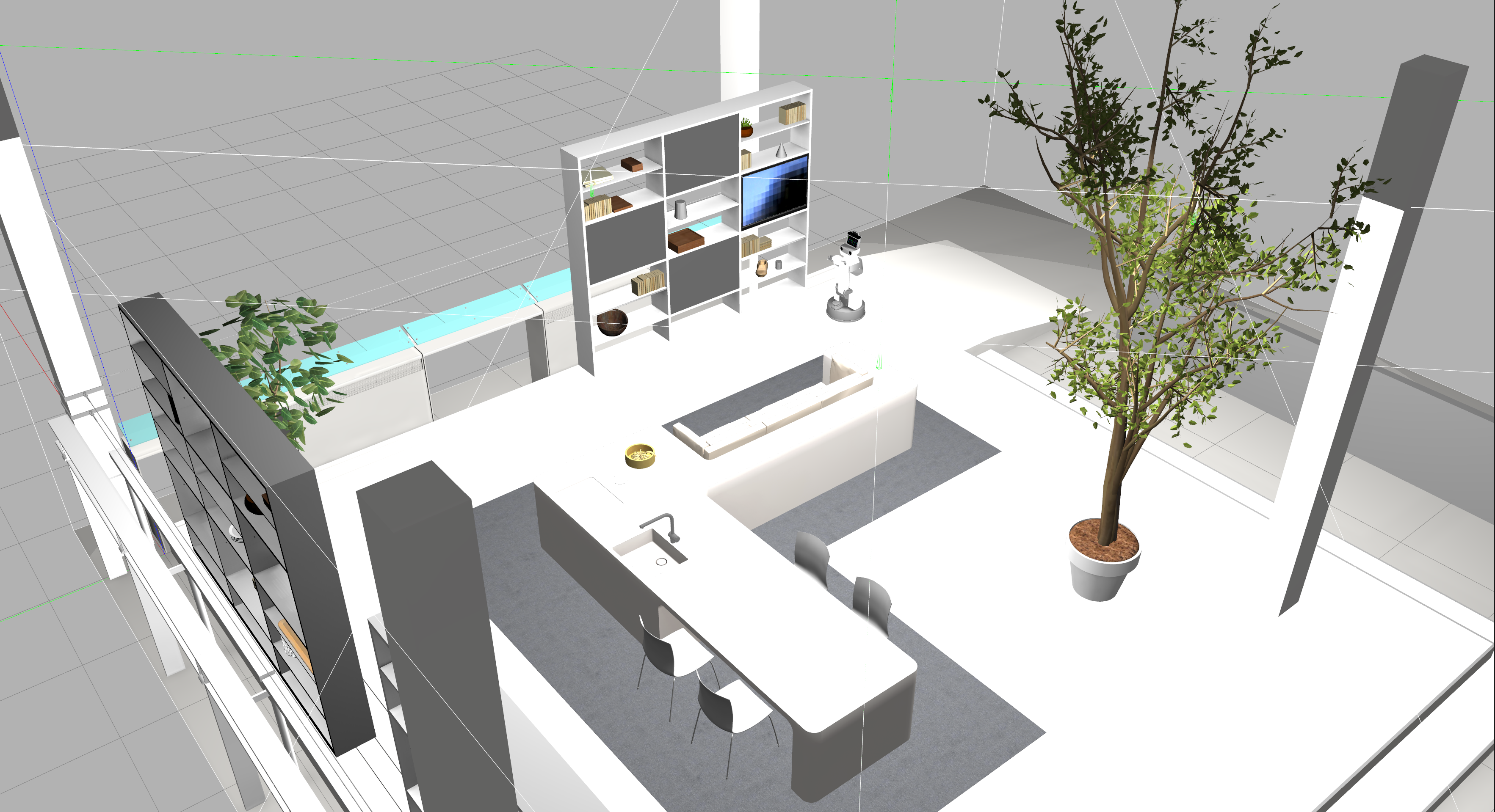}
        \subcaption{Gazebo view as a ground truth.}
        \label{fig:HSRsimulatorGazebo}
    \end{minipage}
    \begin{minipage}{0.45\hsize}
        \centering
        \includegraphics[width=\linewidth]{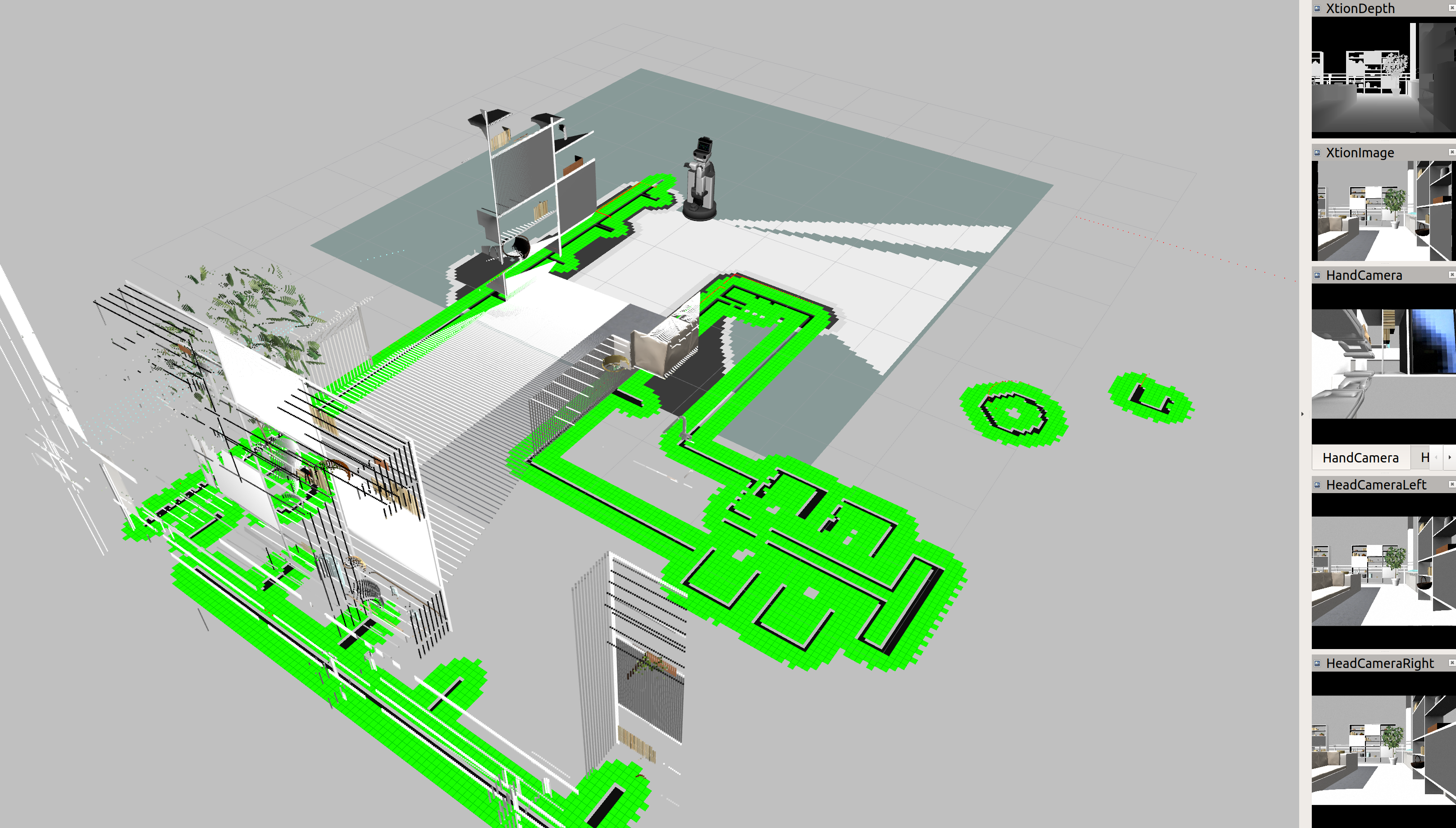}
        \subcaption{Rviz view as a robot perception view.}
        \label{fig:HSRsimulatorRviz}
    \end{minipage}
    \caption{\hsr\ simulator based on \ros\ Gazebo.}
    \label{fig:HSRsimulator}
\end{figure*}

\begin{figure*}[t]
    \centering
    \begin{minipage}{0.48\hsize}
        \centering
        \includegraphics[width=\linewidth]{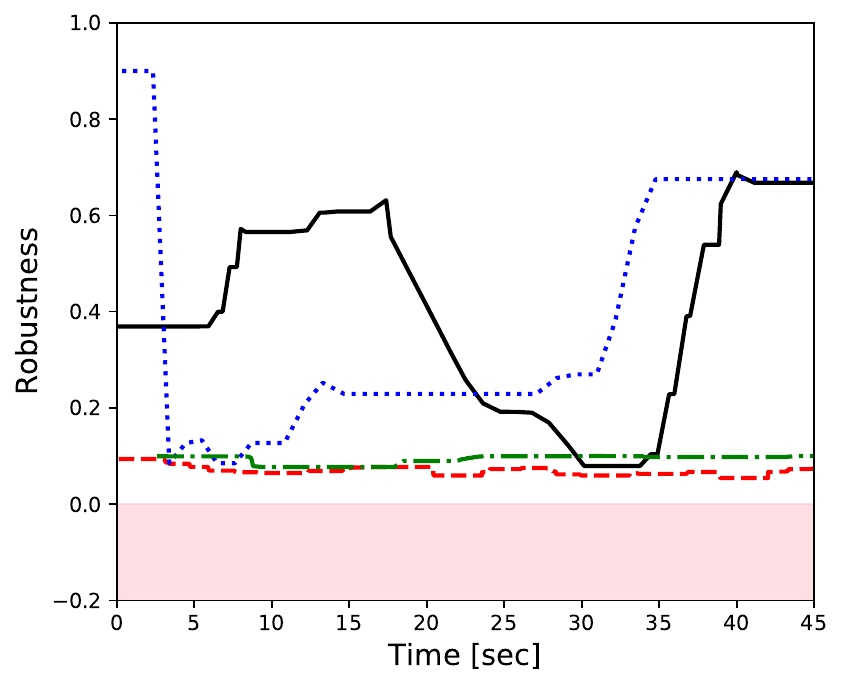}
        \subcaption{The case of all systems satisfying safety specification.
        \url{https://youtu.be/6xFHhv-F9A4}}
        \label{fig:systemGreen}
    \end{minipage}
    \begin{minipage}{0.48\hsize}
        \centering
        \includegraphics[width=\linewidth]{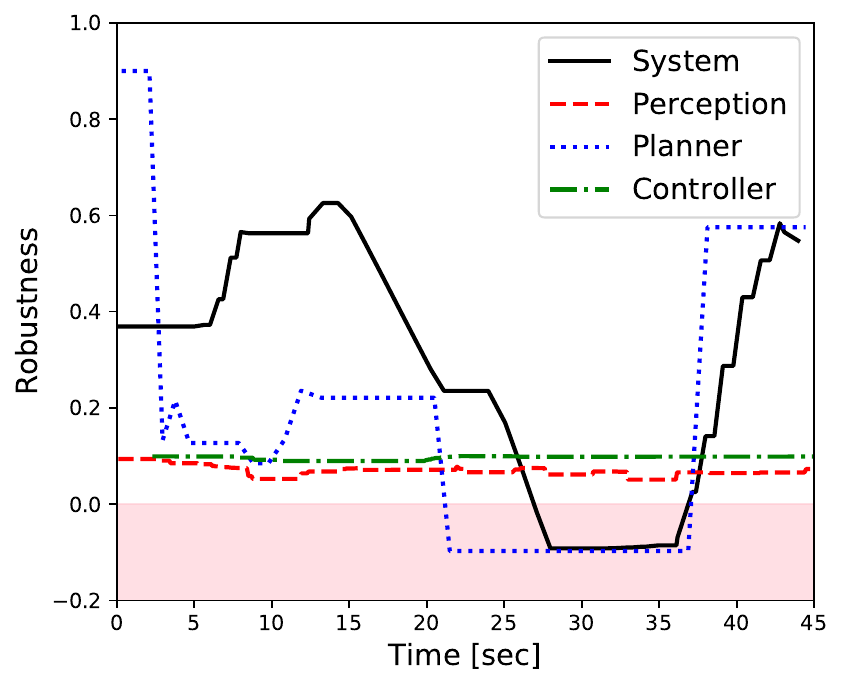}
        \subcaption{The case of violation because of the planner fault. \url{https://youtu.be/nGJ8siD8Fgk}}
        \label{fig:plannerFail}
    \end{minipage}

    \caption{\hsr\ experiment on fault localization with system (Eq.~\ref{eq:systemSpec}) and subsystem safety specifications (Eq.~\ref{eq:perceptionSpec}-\ref{eq:controllerSpec}) in \stl\ with \rtamt: (a) shows all specifications are satisfied (robustness is greater than zero). (b) shows the system specification is finally violated (robustness is below zero) and we can expect the planner subsystem is the cause because it fails before the system since other perception and planner subsystems satisfy the specification.}
    \label{fig:experimentROS}
\end{figure*}

Now we share the results of evaluation of system safety and subsystem level of fault localization when the system has a violation, based on the formal specifications above (Fig.~\ref{fig:experimentROS}).
Each of the specification parameters are 
$\tau=3$~[sec], $\varepsilon=0.3$~[m] in Eq. \ref{eq:systemSpec} $\tau_1=7$, $\tau_2=5$~[sec], $\varepsilon=0.1$~[m] in Eq.~\ref{eq:perceptionSpec},~$\tau=3$~[sec], $\varepsilon=0.2$~[m]~in Eq.~\ref{eq:plannerSpec}, and $\tau_1=7$, $\tau_2=5$~[sec], $\varepsilon=0.1$~[m/s] in Eq.~\ref{eq:controllerSpec}.
Since all properties are \bfstl, in order to evaluate online, we applied pastification for all properties. 
First, the system all green case is shown in Fig.~\ref{fig:systemGreen}, where we see all the specifications on both the system and subsystem levels satisfy the robustness of Eq.~\ref{eq:systemSpec}, \ref{eq:perceptionSpec}, \ref{eq:plannerSpec}, \ref{eq:controllerSpec} all the time.
On the other hand,  Fig.~\ref{fig:plannerFail} shows a violation of system specification and subsystem level of fault localization.
The planner is intentionally injected with a malfunction, the RRT being cut-off and given the wrong way points, causing a collision with a desk.
The experiment shows that planner specification Eq.~\ref{eq:plannerSpec} is violated due to fault injection, so that the system specification in Eq.~\ref{eq:systemSpec} is violated either.
In reality, it is difficult for a designer to detect a system fail based on a subsystem fail because there are many subsystems, and a subsystem failure does not always cause a system failure.
However, we assume a designer can at least detect a suspicious subsystem when the system failure.
This \stl\ assumption-based fault localization approach can also find a system issue caused by perception or controller failure ~\cite{yamaguchi2020specification}.

\subsection{\matlabSimulink\ Model}
\label{sec:experiment_Simulink}
\begin{figure*}[t]
    \centering
	\includegraphics[width=0.8\linewidth]{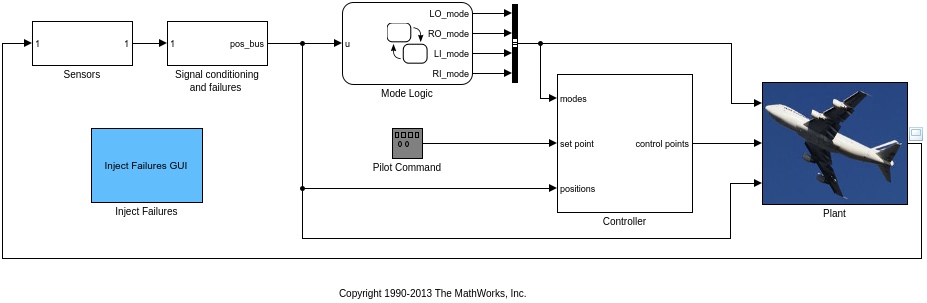}
    \caption{Aircraft Elevator Control System (AECS).}
    \label{fig:aecs}
\end{figure*}
In this section, we show how the \rtamt\ library integrated into \matlabSimulink\ and its resulting robustness monitors can be used to support Model-Based Development (MBD) of \cps.
We explore two scenarios using the aircraft elevator control system (AECS), a model that illustrates MBD of a fault-tolerant control system (Fig.~\ref{fig:aecs}).These flight control systems are typically associated to the horizontal tails of the aircraft. They are responsible for the control of the aircraft's pitch. The system incorporates safety-relevant redundant components, namely: (1) four hydraulic actuators (two per elevator), three hydraulic controllers for driving the actuators, two Primary Flight Control Units (PFCUs) and two control modules for each actuator, one providing full range control law and limited range (degraded) control law.
AECS has one input, the Pilot Command (PC), and two observable outputs, the position of left and right actuators (LEP and REP).
The main requirement intuitively states that the actuators must follow PC within reasonable time and error. 
We translate the requirement for the LEP (the requirement for the REP is similar) to the following \iastl\ specification:

$$
\globally(\uparrow(pc \geq m) \imply \globally_{[0,T]} \finally_{[0,t]} (\lvert pc-lep \rvert \leq n)),
$$
\noindent where $m, n, t$ and $T$ are constants.

\begin{figure*}[t]
    \centering
    \begin{minipage}{0.46\hsize}
        \centering
        \includegraphics[width=\linewidth]{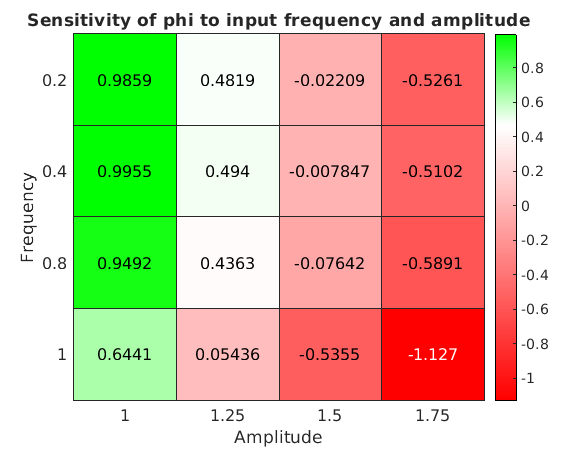}
        \subcaption{Sensitivity analysis.}
        \label{fig:sa}
    \end{minipage}
    \hspace{24pt}
    \begin{minipage}{0.46\hsize}
        \centering
        \includegraphics[width=\linewidth]{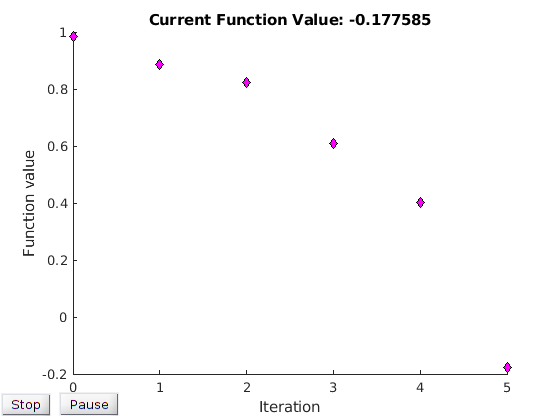}
        \subcaption{Falsification testing.}
        \label{fig:ft}
    \end{minipage}
    \caption{AECS experiment:
        (a) a heat-map of the robustness on the input parameters.
        (b) a gradual reduction of the robustness to falsify the requirement by a global optimizer.}
\end{figure*}
We use \rtamt\ to do {\em sensitivity analysis} and {\em falsification testing}, two common approaches supported by tools like Breach and S-TaLiRo.
Sensitivity of the model robustness to its \stl\ requirements is studied by uniformly varying input parameters, simulating the model for each combination and monitoring the simulation outcomes against the requirements. 
In this case study, we test the system against PC step signals. A step signal is fully characterized by two parameters, the frequency and the amplitude of individual steps in the signal. We varied the frequency of the step input between $0.2$ and $1$ and the amplitude between $1$ and $1.75$, generating in total $16$ simulations, one for each (amplitude, frequency) concrete pair. For every simulation, we evaluated how robust is the system behavior with respect to the bounded stabilization requirement for that specific input signal.
Fig.~\ref{fig:sa} shows a heat-map visualizing the outcomes. 
This graphical representation of the results helps understanding how different input parameters affect the requirements and enable identifying critical parameter regions.

Given an \stl\ specification and the system model, falsification testing aims to identify an input signal that results in the violation of the requirement. It does so by framing the test generation as a global optimization problem. 
The global optimizer (used as an off-the-shelf black-box component in this context) has the task of minimizing the system robustness to a requirement by finding appropriate inputs that steer the system in that direction, where the quantitative output of the monitor is used as the objective function. 
In essence, the global optimizer performs an iterative search in the space of input signals, where each iteration aims at steering the system towards lower robustness. 
The outcome is shown in Fig.~\ref{fig:ft} -- every point $(x,y)$ corresponds to a simulation of the system where $x$ is the iteration number and $y$ is the robustness of the system behavior to the requirement as measured by the monitor.

\section{Related Work}
\label{sec:relatedWork}

Linear Temporal Logic (\ltl)~\cite{pnueli1977temporal} is a well-known formal specification language for describing temporal properties of reactive systems.   
Signal Temporal Logic (\stl)~\cite{mn04} is an extension of \ltl\ that allows reasoning about real-valued signals and their real-time properties. Originally, both \ltl\ and \stl\ are interpreted using \emph{qualitative} semantics -- an observed behavior either satisfies or violates its specification. 
Fainekos and Pappas~\cite{fainekos,fainekos-robust} first proposed equipping temporal logic with \emph{spatial} quantitative semantics, based on the \emph{infinity norm}. 
Donz\'{e} and Maler's~\cite{robust1} adapt spatial robustness to \stl\ and extend it with a complementary notion of \emph{time robustness}. Intuitively, time robustness indicates how much an observed signal needs to be shifted in time in order to satisfy or violate the specification.
Ferr\`{e}re et al. developed an efficient algorithm for quantitative evaluation of \stl~\cite{robust2}, rendering the approach practical.

A combined notion of time and space robustness is proposed in~\cite{AbbasMF14}. Akazaki extended \stl\ with averaged temporal operators in~\cite{avstl}, which allow to quantify \emph{how often} a temporal operator is satisfied within a bounded interval.

Evaluating temporal specifications with quantitative semantics was implemented in S-TaLiRo~\cite{staliro} and Breach~\cite{breach} tools. These works established the theoretical background for our library, which implements \stl\ with infinity-norm quantitative semantics.
 
Runtime verification of declarative specifications has been extensively studied. 
Havelund and Rosu~\cite{safety-monitors} proposed a dynamic programming algorithm to monitor past-time Linear Temporal Logic (ptLTL) specifications. 
Our inspiration to strive for a simple and elegant online monitoring procedure partly arose from this seminal paper. 
Reinbacher et al. propose in~\cite{Reinbacher2013,Reinbacher2014a} synthesizable hardware monitors from past-time Metric Temporal Logic (ptMTL) interpreted over discrete-time. 
Similarly, Jak\v{s}i\'{c} et al. developed monitors that could be implemented in hardware for bounded-future Signal Temporal Logic (\bfstl). 
Reinbacher et al.~\cite{reinbacher2014temporal} proposed monitoring procedures for Mission-time Linear Temporal Logic (MLTL), a variant of discrete-time MTL. 
The monitoring procedure consists of two types of observers: a synchronous one that yields a $3$-valued instant abstraction of the satisfaction check, and an asynchronous one that makes this abstraction concrete at a later (bounded) time.
This approach also allows a probabilistic estimation of the system health using a Bayesian network on top of the synchronous observers. 
This monitoring procedure was implemented in the R2U2 tool~\cite{r2u2-1,r2u2-2}. 
More recently, MLTLM~\cite{DBLP:conf/cav/HariharanKWJR22} was proposed as a
language that allows to define temporal specifications over variables of multiple type.
Our library supports quantitative semantics and makes a distinction between inputs and outputs.

Another relevant field of research is \emph{stream runtime verification (SRV)}, in which runtime monitors are specified by describing operations that transform input streams of data to output streams of data. 
Precursor ideas of collecting statistics over execution traces can be found in~\cite{DBLP:journals/entcs/FinkbeinerSS02}. 
Lola~\cite{DBLP:conf/time/DAngeloSSRFSMM05} proposes a SRV approach for synchronous systems. Lola 2.0~\cite{faymonville2016stream} extends the original language with parameterization and multiple temporal time scales. 
RTLola~\cite{faymonville2017real} and Tessla~\cite{convent2018tessla} provide support for specifying and monitoring real-time properties of reactive systems. 
Finally, Striver~\cite{gorostiaga2018striver} provides a more general language that enables expressing other real-time monitoring languages. 
SRV approaches are powerful and typically more general than logic-based approaches. Temporal logic formalism, such as LTL and STL, can be often expressed with SRV. 
However, this comes at the price of requiring the user to explicitly encode the temporal logic semantics and thus losing a useful layer of abstraction between the user and the underlying formalism.

The problem of online robustness monitoring was studied in~\cite{DokhanchiHF14,DeshmukhDGJJS17}. 
Dokhanchi et al.~\cite{DokhanchiHF14} proposed an online monitoring approach which required a model of the system to make predictions about its behavior. 
In~\cite{DeshmukhDGJJS17}, Deshmukh et al. 
proposed an interval-based approach of online evaluation that allows estimating the minimum and the maximum robustness with respect to both the observed prefix and unobserved trace suffix. 
An alternative approach to online monitoring proposed a specification language that relaxes the causality restriction and allows the output to depend on a bounded amount of future input~\cite{mamouras20}. 
Algebraic approaches to runtime verification of temporal logic equipped with quantitative semantics were studied in~\cite{arv} and~\cite{mamouras21}.
These papers considered alternative ways to address online (qualitative and quantitative) monitoring of temporal logic specifications.

Pattern matching of Timed Regular Expressions (TREs)~\cite{tre1,tre2} has been a lively area of research in recent years.
Ulus et al.~\cite{patterns} proposed an offline pattern matching procedure for TRE with qualitative semantics and interpreted over continuous time, utilizing a novel matching approach that used operations on $2$-dimensional zones. 
Ulus et al.~\cite{dogan-online} proposed an online variant of the procedure, also using the idea of derivatives adapted to the continuous-time.
These pattern matching algorithms were implemented in the MONTRE tool~\cite{dogan-tool}. 
Unlike previous automata-based matching algorithms for TREs which have been developed in~\cite{ichiro-moore,ichiro-skipping,ichiro-tool}, our work is centered on temporal logics rather than regular expressions.

\section{Future Work and Conclusions}
\label{sec:conclusion}
We presented \rtamt, a library for generating online and offline monitors from declarative specifications.
We have integrated \rtamt\ into \ros\ as \rosrtamt\ and applied it to robotic applications as a robotics domain experiment.
We have also integrated \rtamt\ into \matlabSimulink\ and applied it to AECS as a control domain experiment.
The tool provides a flexible and modular library that enables practitioner specific specification language implementation.

In order to support process management that handles both time and event driven cases, we assumed dense-time as a perfect continuous clock, a realistic assumption in many applications.
As a consequence, we have extended \rtamt\ with an event-driven online monitoring algorithm for \bfstl\ in which sensor measurements will be allowed to arrive at any point on the dense-time axis.

The current library infrastructure already allows the monitor to continuously publish the computed robustness degree. 
This will enable decentralizing and distributing monitors. 
We will investigate both the theory and the implementation of distributed \rtamt\ monitors, especially in the context of robotic applications.

We plan to provide support for additional specification languages, but also for other semantic extensions of existing languages such as \stl\ with weighted edit distance semantics.

Finally, we will further evaluate the library and the tool in other application domains. 
Actually, online monitoring of declarative specifications has various potential use-cases in the broad CPS domain and not only in robotics.
Typical controller properties are naturally expressed in specification languages such as \stl\ \cite{kapinski2016st}.
For Advanced Driver Assistance Systems (ADAS), there has been work on defining critical scenarios for precrash safety systems~\cite{najm2007pre}.
For self-driving systems, Responsibility-Sensitive Safety (RSS)~\cite{koopman2019autonomous} has been proposed, based on a clear definition by a math formula.
Since RSS is not a temporal logic-based definition, it would be interesting to encode it in \stl~\cite{hekmatnejad2019encoding}.

We plan to evaluate our library in such scenarios, both in a simulation environment and during actual in-field physical testing.
Search-based testing~\cite{breach, staliro} also uses formalisms such as \stl/MTL as a cost function to find input vectors that steer the system to property violation.
This approach defines test generation as an optimization problem.
In each iteration, the simulator executes the new input vector and the outcomes are evaluated in an offline fashion. 
We plan to explore whether \rtamt\ can make this process more efficient and dynamic by evaluating the outputs after each simulation step, adapting online search.
VerifAI~\cite{dreossi2019verifai} is a simulation-based verification environment for AI-based applications that includes search-based testing which is implemented in \python\ and is one  candidate for the application of \rtamt. 
Finally, we will consider potential applications of \rtamt\ in the verification and validation of perception systems based on neural networks.
Several formal verification targets~\cite{hekmatnejad2019encoding, rong2020lgsvl, vitelli2021safetynet}, approaches~\cite{liu2019algorithms, tuncali2018simulation, date2020application, ghosh2021counterexample}, and specifications~\cite{dreossi2019formalization} have been recently proposed.
Unexpected failure of such perception systems can be checked during operation using our library.

This version of the article has been accepted for
publication, after peer review (when applicable) but is not the Version of Record and does not reflect postacceptance improvements, or any corrections. The Version of Record is available online at:
\url{http://dx.doi.org/10.1007/S10009-023-00720-3}. Use of this Accepted Version is subject to the publisher’s Accepted Manuscript
terms of use \url{https://www.springernature.com/gp/open-research/policies/accepted-manuscript-terms}.

\bibliography{main.bib}



\end{document}